\documentclass[%
 reprint,
superscriptaddress,
 amsmath,amssymb,
 aps,
pra,
]{revtex4-2} 
\usepackage{amsmath}
\usepackage{amssymb}
\usepackage{graphicx}
\usepackage{bm}
\usepackage{slashed}
\usepackage{xcolor}
\usepackage[colorlinks=true,urlcolor=blue,citecolor=blue]{hyperref}

\begin{document}

\preprint{APS/123-QED}

\title{Correlated decoherence in a common environment activated by relative motion}

\author{Yang Wang}
\affiliation{School of Information Science and Engineering, Shandong Institute of Petroleum and Chemical Technology, Dongying, 257061, China.}

\author{Zhilei Sun}
\affiliation{School of Information Science and Engineering, Shandong Institute of Petroleum and Chemical Technology, Dongying, 257061, China.}

\author{Feiyi Liu}
\email{fyliu@cxtc.edu.cn}
\affiliation{School of Physics, Electrical and Energy Engineering, Chuxiong Normal College, Chuxiong 675000, China.}

\author{Min Guo}
\affiliation{School of Physics, Electrical and Energy Engineering, Chuxiong Normal College, Chuxiong 675000, China.}

\author{Yuhan Jiang}
\affiliation{School of Physics, Electrical and Energy Engineering, Chuxiong Normal College, Chuxiong 675000, China.}
 
 \author{Mingyang Liu}
\affiliation{School of Physics, Electrical and Energy Engineering, Chuxiong Normal College, Chuxiong 675000, China.}

\date{\today}

\begin{abstract}

We study two spatially separated boundary subsystems coupled to a common structured environment under relative motion within a Gaussian open-system framework. By integrating out the environment, we obtain an influence functional governed by a dressed environmental correlator evaluated at the boundary positions. Its retarded and Hadamard components govern coherent mediation and correlated fluctuations, respectively. Relative motion activates the leading resonant contribution to correlated decoherence through Doppler-shifted spectral overlap of the boundary excitations. For identical boundary dispersions, this contribution has an ideal narrow-linewidth onset at $v>2u_\phi$. Below this value, no resonant momentum shell exists, whereas above it a finite shell opens and produces an enhanced correlated-decoherence contribution. Finite spectral linewidth generates subthreshold tails and rounds the ideal onset into a crossover centered on the same kinematic condition. The leading resonant contributions to motion-induced excitation production and correlated decoherence probe the same on-shell environmental structure. The proposed readout of this two-subsystem reduced-dynamics signature uses either differential-coherence measurements or a two-node noise cross-spectrum.

\end{abstract}


\maketitle

\section{Introduction}
\label{Introduction}

The interaction between quantum systems and their environments plays a central
role in decoherence, dissipation, and the quantum-to-classical transition
\cite{Paz2002,Schlosshauer2007,RevModPhys.75.715,PhysRevLett.104.200401,doi:10.1142/S0129055X03001631}.
A microscopic description of reduced dynamics is naturally provided by the
influence-functional approach, beginning with the Feynman--Vernon formalism
and its extensions to quantum dissipation and Brownian motion
\cite{Feynman1963,Caldeira1983,Leggett1987,Zurek1991vd}. These ideas were
further developed in nonequilibrium quantum field theory, where reduced
dynamics can be derived from first principles using closed-time-path (CTP)
techniques and path-integral methods
\cite{Hu1992,Ingold2002,10.1063/1.531046}. More broadly, open quantum system
(OQS) frameworks provide a unified language for dissipation, noise, and
decoherence across a wide range of platforms
\cite{Breuer2002,Weiss2012,Rivas2012,banerjee2018open}. Closely related
Keldysh and functional-integral methods have also become standard tools for
driven-dissipative systems, nonequilibrium phase transitions, quantum
many-body dynamics, and non-Markovian trajectories
\cite{Sieberer2016,Altland2010,Berges2004,Strunz1999}.

Within this OQS setting, a central issue is the role of common and structured
environments in shaping the reduced dynamics. A shared bath can generate
collective decoherence and correlations between otherwise independent
subsystems \cite{Braun2002}, and can also mediate entanglement generation and
nontrivial dynamical correlations in open systems
\cite{Reina2002,McCutcheon2011}. The impact of environmental structure has
been explored in studies of structured spectral densities \cite{Chin2013},
non-Markovian decoherence and memory effects
\cite{PhysRevA.94.042110,DeVega2017,Shrikant2023}, coherence generation and
transfer in correlated environments
\cite{Sarkar97h,PhysRevA.106.032220,PhysRevResearch.7.L012068,PhysRevA.101.013822,Wu4ltm},
and anomalous steady-state or memory-driven coherence phenomena
\cite{PhysRevA.110.052220,PhysRevA.105.012209,zvkls7hy}. In quantum optics and
mesoscopic physics, environmental fluctuations have long been understood to
appear directly as measurable noise and decoherence channels
\cite{Carmichael1999,Rotter_2015,RevModPhys.88.041001,1609231253819,Clerk2010,Rivas2012}.
In superconducting and other quantum devices, studies of electronic noise,
measurement backaction, engineered dissipation, and correlated fluctuations
have further highlighted the role of environmental correlations in realistic
decoherence processes
\cite{10.1063/5.0197142,PhysRevApplied.20.034038,Liu_2022,harrington2022,Campbell_2026,Schuetz2015,vonLupke2020}.
Spatially and temporally correlated noise has likewise been emphasized in
realistic devices and in structured many-body environments
\cite{Strathearn2018,Tamascelli2019,Kirton2020,Chenu2020,sung2019non,PhysRevResearch.7.023073,Zou2024SpatiallyCorrelatedNoise,Brattegard2024,Brattegard2026, PhysRevResearch.6.043222}.

A complementary line of research concerns motion-induced nonequilibrium
phenomena. Quantum friction and noncontact dissipation were formulated early
for moving bodies and fluctuating fields
\cite{Pendry1997,Volokitin2007}. Functional approaches to moving
dispersive or imperfect boundaries encode internal material degrees of
freedom in nonlocal actions used to calculate Casimir energies. The same
methods describe imaginary parts of in-out effective actions, dissipative
forces, and Schwinger--Keldysh noise kernels
\cite{Fosco2007,Fosco2008,Fosco2011}. Later work analyzed nonequilibrium
fluctuations, vacuum-excitation effects, and engineered motion in greater
detail
\cite{Maghrebi2015,Intravaia2014,Millen2020,10.1063/5.0083067,z3gm32jn,PhysRevA.106.052205,fskmy179}.
The anomalous-Doppler interpretation of motion-activated excitation
channels is particularly explicit in quantum \v{C}erenkov radiation
\cite{Maghrebi2013}. Motion-induced dissipation and decoherence have also
been studied more directly in moving particle-surface settings, including
CTP treatments of decoherence and its thermal corrections
\cite{Farias2016,Viotti2019}. Related single-probe observables include
motion-dependent local decoherence and velocity-dependent
corrections to an accumulated geometric phase \cite{Viotti2021, Farias2020}. These proposals complement the two-subsystem cross-noise observable
analyzed in the present study.

In parallel, spatially correlated decoherence and common-bath noise have been
formulated within reduced-dynamics approaches \cite{Jeske2013}. More recent
work has sharpened several adjacent aspects of the problem, including driven
qubits under spatially correlated noise
\cite{Zou2024SpatiallyCorrelatedNoise}, mobile-qubit coherence under shuttling
or motion-related noise filtering \cite{Krzywda2026MobileSpinQubits}, driven
particle--bath open-system formalisms \cite{Gamba2025DrivenParticleBath}, and
nontrivial two-body dynamics induced by correlated environments
\cite{Tyagi2024HuygensClock}. Collectively, these studies show that relative
motion can open dissipative channels through nonequilibrium spectral matching
and environmental excitation production, while correlated environmental
noise can strongly reshape reduced dynamics. The role of these two
ingredients in the reduced dynamics of spatially separated subsystems coupled
to a common structured environment remains less clear. This motivates a
reduced-dynamics treatment that distinguishes the full common-environment
noise kernel from its motion-activated resonant contribution.

To address this gap, we apply an established Gaussian OQS framework
based on the influence-functional and CTP formalisms. We show that relative motion Doppler-shifts the boundary spectra.
Their overlap activates the leading resonant contribution to correlated
common-bath decoherence. The onset
depends on the boundary dispersions. For identical branches, its ideal
zero-linewidth condition is $v>2u_\phi$. The full cross-noise kernel generally
remains finite below this value, while finite linewidth rounds the onset into
a crossover. In the reduced dynamics, coherent mediation and correlated noise are governed
by different components of the same dressed environmental correlator.
The leading resonant parts of excitation production and correlated
decoherence probe the same on-shell environmental structure. The cross-noise
contribution provides an experimentally accessible two-body differential
signature. We also discuss a proposed proof-of-principle synthetic-motion
superconducting--phononic emulator and qualitative connections to cold-atom,
graphene, and plasmonic systems.

The remainder of this paper is organized as follows. Section~\ref{model}
introduces the model and derives the effective environmental kernel and
dressed propagator. Section~\ref{imag} identifies the motion-activated
excitation channel, while Sections~\ref{reduced_dynamics} and
\ref{decoherence_connection} develop the reduced dynamics and its connection
to decoherence. Section~\ref{comparison} compares different environmental
phases, Section~\ref{decoherence_rate} analyzes the decoherence rate, and
Section~\ref{experimental_relevance} discusses experimental relevance and
possible implementations. Section~\ref{conclusion} concludes the paper.


\section{Model and effective environmental description}\label{model}
\subsection{Physical setup and subsystem--environment partition}\label{setup}

\begin{figure}[t]
    \centering
    \includegraphics[width=1.1\linewidth]{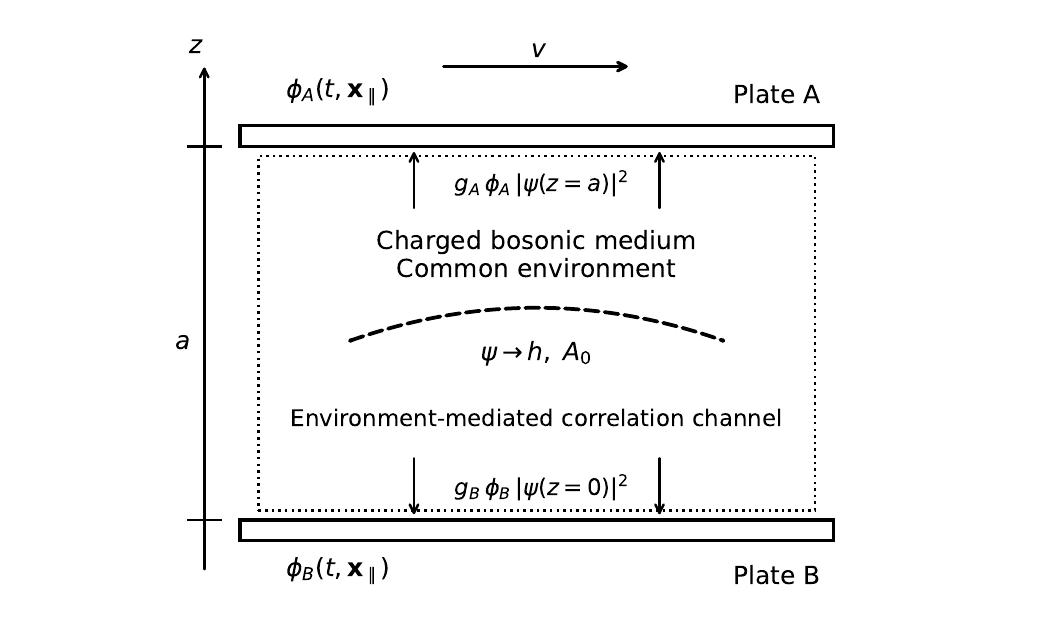}
\caption{Schematic of the setup. Plate $B$ is fixed at $z=0$, while plate $A$ at $z=a$ moves with velocity $v$ along $x$.
The region between the plates contains a bosonic medium that acts as a common environment for the boundary modes $\phi_A$ and $\phi_B$.
The dominant low-energy environmental channel is represented by the amplitude fluctuation mode $h$.}
\label{setup_model}
\end{figure}

We consider two parallel planar boundaries separated by a distance $a$. Plate $A$ is located at $z=a$ and moves with a constant velocity $v$ along the $x$ direction, while plate $B$ is fixed at $z=0$. The region $0<z<a$ is filled with a charged bosonic medium coupled to an Abelian $U(1)$ gauge field. Within the OQS framework, the effective boundary modes define the subsystem, while the cavity medium and its screened gauge response act as a common environment. Tracing out the cavity degrees of freedom produces the reduced boundary dynamics, including both coherent inter-boundary coupling and dissipative effects. Figure~\ref{setup_model} shows the geometry of the setup, the subsystem-environment partition, and the boundary-medium couplings used in the effective description.

The total action is written as
\begin{equation}
S = S_{\rm sys} + S_{\rm env} + S_{\rm EM} + S_{\rm int},
\label{S_total_model}
\end{equation}
where $S_{\rm sys}$ describes the boundary subsystem, $S_{\rm env}$ the charged bosonic medium inside the cavity, $S_{\rm EM}$ the gauge sector, and $S_{\rm int}$ the boundary--environment coupling. Throughout this work we use natural units $\hbar=c=1$.

Each plate supports a real scalar boundary field, denoted by $\phi_A(t,\mathbf{x}_\parallel)$ and $\phi_B(t,\mathbf{x}_\parallel)$, where $\mathbf{x}_\parallel=(x,y)$ labels the coordinates parallel to the interfaces. These fields represent effective surface excitations confined to the plates. Their free action is taken to be
\begin{equation}
\begin{aligned}
S_{\rm sys}
=&\frac12 \int dt\, d^2\mathbf{x}_\parallel
\Big[
\frac{1}{u_\phi^2}(\partial_t\phi_A)^2
- |\nabla_\parallel \phi_A|^2
- m_\phi^2 u_\phi^2 \phi_A^2\\
&+\frac{1}{u_\phi^2}(\partial_t\phi_B)^2
- |\nabla_\parallel \phi_B|^2
- m_\phi^2 u_\phi^2 \phi_B^2
\Big].
\end{aligned}
\label{S_sys}
\end{equation}
Here $u_\phi$ is the propagation velocity of the boundary modes and $m_\phi$ is an effective mass parameter. Fourier transforming along the boundary directions gives the dispersion relation
\begin{equation}
\Omega_{\mathbf{k}}=u_\phi \sqrt{k_\parallel^2 + m_\phi^2 u_\phi^2},
\qquad
k_\parallel \equiv |\mathbf{k}|=\sqrt{k_x^2+k_y^2}.
\label{boundary_dispersion}
\end{equation}

The cavity medium is modeled by a charged complex scalar field $\psi$ confined to $0<z<a$ and minimally coupled to an Abelian gauge field $A_\mu=(A_0,\mathbf{A})$. Its dynamics are described by
\begin{equation}
\begin{aligned}
S_{\rm env}
&=\int dt\, d^2\mathbf{x}_\parallel\, dz\;
\Theta(z)\Theta(a-z)\\
\Bigg\{
&\frac{1}{u_\psi^2}
\big|(\partial_t - i\mu - i e A_0)\psi\big|^2-\big|(\nabla - i e \mathbf{A})\psi\big|^2\\
&-m_\psi^2 u_\psi^2 |\psi|^2-\frac{\lambda_\psi}{4}|\psi|^4
\Bigg\},
\end{aligned}
\label{S_env_complex}
\end{equation}
while the gauge field itself is governed by the Maxwell action
\begin{equation}
S_{\rm EM}=-\frac14\int dt\, d^3x\; F_{\mu\nu}F^{\mu\nu},
\label{S_EM}
\end{equation}
with
\begin{equation}
F_{\mu\nu}=\partial_\mu A_\nu-\partial_\nu A_\mu.
\end{equation}
Here $\mu$, $e$, $m_\psi$, and $\lambda_\psi$ denote the chemical potential, gauge charge, bare scalar mass parameter, and quartic self-interaction strength, respectively, and $u_\psi$ is the characteristic propagation velocity of the scalar medium.


\subsection{System--environment coupling}\label{coupling}

At finite density, expanding around a homogeneous stationary background requires some care in a gauged theory. A nonzero chemical potential must be accompanied by a neutralizing static background to satisfy Gauss-law consistency at the saddle point  \cite{Elze1987,Davoudi2005}. We therefore write
\begin{equation}
A_0=\bar A_0+\delta A_0,
\qquad
\mu_{\rm eff}\equiv \mu+e\bar A_0,
\label{A0_background_split}
\end{equation}
and choose the background so that the action contains no term linear in $\delta A_0$. For notational simplicity, we henceforth relabel $\delta A_0\to A_0$ and $\mu_{\rm eff}\to\mu$. This should be understood as an effective neutralized-medium description rather than a microscopic treatment of the full background-charge sector.

The boundary subsystem couples locally to the density of the cavity medium at the two interfaces. This choice is motivated by the low-energy regime of interest, in which we focus on the dominant quasi-static density-response channel of the cavity medium rather than on the full set of electromagnetic retardation effects. We therefore model the interaction as
\begin{equation}
\begin{aligned}
S_{\rm int}
=&\int dt\, d^2\mathbf{x}_\parallel\bigg[g_A\,\phi_A(t,\mathbf{x}_\parallel)\,\big|\psi\!\left(t,x+vt,y,a\right)\big|^2\\
&+g_B\,\phi_B(t,\mathbf{x}_\parallel)\,\big|\psi\!\left(t,\mathbf{x}_\parallel,0\right)\big|^2\bigg],
\label{S_int_density}
\end{aligned}
\end{equation}
where $g_A$ and $g_B$ are effective boundary couplings. Here $\mathbf{x}_\parallel=(x,y)$ denotes the comoving coordinate on plate $A$ and the laboratory coordinate on plate $B$, so that the moving-boundary coupling later gives rise to the Doppler combination $\omega-vk_x$. Eq.~\eqref{S_int_density} shows that both boundary modes couple to the same bulk density operator at opposite sides of the cavity. As a result, the medium acts as a common bath and generates both local and cross kernels in the reduced dynamics. The resulting theory is thus an effective low-frequency, near-field description rather than a complete treatment of radiative processes.

To make the low-energy environmental modes explicit, it is convenient to parameterize the complex scalar field in polar form,
\begin{equation}
\psi(x)=\frac{1}{\sqrt{2}}\,\rho(x)e^{i\theta(x)},
\qquad
|\psi(x)|^2=\frac{\rho^2(x)}{2}.
\label{polar_decomposition}
\end{equation}
In the condensed phase analyzed in the next section, we expand around a nonzero stationary amplitude $\rho_0$ by writing
\begin{equation}
\rho=\rho_0+h,
\label{rho_expand}
\end{equation}
where $h$ denotes the amplitude fluctuation. With this decomposition, the boundary interaction in Eq.~\eqref{S_int_density} becomes
\begin{equation}
\begin{aligned}
S_{\rm int}=\int dt\, d^2\mathbf{x}_\parallel
\Big[
&\frac{g_A\rho_0^2}{2}\phi_A+g_A\rho_0\,\phi_A\,h\!\left(t,x+vt,y,a\right)\\
&+\frac{g_A}{2}\phi_A\,h^2\!\left(t,x+vt,y,a\right)\\
&+\frac{g_B\rho_0^2}{2}\phi_B+
g_B\rho_0\,\phi_B\,h\big(t,\mathbf{x}_\parallel,0\big)\\
&+\frac{g_B}{2}\phi_B\,h^2\big(t,\mathbf{x}_\parallel,0\big)\Big].
\end{aligned}
\label{Sint_expanded}
\end{equation}
The terms proportional to $\rho_0^2\phi_{A,B}$ merely shift the equilibrium configuration of the boundary fields and can be absorbed into local counterterms or into a redefinition of the background boundary configuration. Within the Gaussian weak-fluctuation approximation adopted later, the cubic terms $\phi_{A,B}h^2$ are neglected, and the leading system--environment coupling reduces to the bilinear form
\begin{equation}
\begin{aligned}
S_{\rm int}^{(1)}=\int dt\, d^2\mathbf{x}_\parallel\Big[&\lambda_A\,\phi_A(t,\mathbf{x}_\parallel)\,h\!\left(t,x+vt,y,a\right)\\
&+\lambda_B\,\phi_B(t,\mathbf{x}_\parallel)\,h\!\left(t,\mathbf{x}_\parallel,0\right)\Big],
\end{aligned}
\label{S_int_linear}
\end{equation}
where $\lambda_A\equiv g_A\rho_0$ and $\lambda_B\equiv g_B\rho_0$. This linearized coupling will be the starting point for the influence-functional treatment.

Because the medium is gauged, the phase variable $\theta$ is not retained below as an independent low-energy scalar excitation in the effective condensed description. Instead, after gauge fixing, or equivalently after passing to gauge-invariant variables, the relevant environmental sector is described by the amplitude fluctuation $h$ together with screened gauge fluctuations \cite{Anderson1963,Gold2002}. In the low-frequency regime considered here, the dominant gauge-mediated contribution comes from the temporal component $A_0$, which captures the density-response channel of the medium. We therefore work in the regime
\begin{equation}
v\ll 1,
\qquad
\omega \ll m_A,
\label{model_approx_window}
\end{equation}
so that the quasi-static density-response channel dominates over transverse retardation. Here $m_A$ denotes the screening scale generated in the condensed medium. For compatibility with the threshold condition $v>2u_\phi$, we also assume $u_\phi\ll 1$, so that a finite velocity window satisfies both constraints. The present description thus relies on the Gaussian weak-fluctuation approximation together with the linearized coupling in Eq.~\eqref{S_int_linear}. Beyond this regime, higher-order environmental correlations and nonlinear interaction terms may generate additional decoherence channels.


\subsection{Effective environmental kernel}\label{kernel}
Having established the coupling structure, we now integrate out the environmental degrees of freedom to obtain the nonlocal kernel governing the reduced boundary dynamics. Within the Gaussian weak-fluctuation approximation, in which the environment is described by a quadratic action and the system--environment coupling is truncated to the bilinear form in Eq.~\eqref{S_int_linear}, this procedure can be carried out exactly. The kernel captures both coherent inter-boundary coupling and, in the closed-time-path (CTP) formulation introduced later, the correlated fluctuations responsible for decoherence.

We construct the dressed propagator of the environmental amplitude mode $h$. At quadratic order, it couples linearly to the temporal gauge component $A_0$, so the relevant quadratic action is
\begin{equation}
S_{A_0}=\int dt\, d^2\mathbf{x}_\parallel\, dz\,\left[\frac12(\nabla A_0)^2+\frac12 m_A^2(z) A_0^2+C\, h A_0\right],
\label{S_A0}
\end{equation}
where
\begin{equation}
C \equiv \frac{2\mu e\rho_0}{u_\psi^2},
\qquad
m_A^2(z)=m_A^2\,\Theta(z)\Theta(a-z).
\end{equation}
Because the action is Gaussian in $A_0$, integrating out $A_0$ yields the induced nonlocal contribution to the effective action of $h$,
\begin{equation}
\begin{aligned}
S_h^{\rm eff}=&\frac12 \int d^4x\,d^4y\,h(x)\,K_h(x,y)\,h(y)\\
&-\frac{C^2}{2}\int d^4x\,d^4y\,h(x)\,D_{00}(x,y)\,h(y),
\end{aligned}
\label{Sh_eff_formal}
\end{equation}
where $K_h$ is the bare quadratic kernel of the amplitude mode and $D_{00}$ is the electrostatic Green function associated with the operator in Eq.~\eqref{S_A0}. The second term represents the nonlocal self-energy induced by screened density fluctuations of the environment.

To evaluate this kernel, it is convenient to work in a mixed representation, Fourier transforming in $(t,\mathbf{x}_\parallel)$ while keeping $z$ explicit. In the quasi-static regime $\omega \ll m_A$, the electrostatic kernel is frequency independent at leading order and satisfies
\begin{equation}
\left[
-\partial_z^2 + k_\parallel^2 + m_A^2(z)
\right]
D_{00}(\mathbf{k};z,z')=\delta(z-z').
\label{D00_eq}
\end{equation}
In the bulk approximation, where the screening is treated as uniform inside the cavity and boundary reflections are neglected at leading exponential order, the solution is
\begin{equation}
D_{00}^{\rm bulk}(\mathbf{k};z,z')=\frac{1}{2\kappa}\,e^{-\kappa|z-z'|},
\label{D00_bulk}
\end{equation}
with
\begin{equation}
\kappa=\sqrt{k_\parallel^2 + m_A^2}.
\end{equation}
The dressed retarded propagator of the amplitude mode then satisfies the Dyson equation
\begin{equation}
\begin{aligned}
&\int_0^a dz''\,\Big[K_h^R(\omega,\mathbf{k};z,z'')-C^2 D_{00}(\mathbf{k};z,z'')\Big]\\
&\times G_h^R(\omega,\mathbf{k};z'',z')=\delta(z-z').
\end{aligned}
\label{Gh_dressed_eq}
\end{equation}
Fourier transforming further in $z$ gives
\begin{equation}
\widetilde G_h^R(\omega,\mathbf{k},q)=\frac{1}{q^2 + \Delta_h(\omega,\mathbf{k}) - \dfrac{C^2}{q^2+\kappa^2}},
\label{Gh_bulk_q}
\end{equation}
with
\begin{equation}
\Delta_h(\omega,\mathbf{k})=k_\parallel^2 + m_H^2 - \frac{\omega^2}{u_\psi^2}.
\label{Delta_h}
\end{equation}
The poles of the propagator satisfy
\begin{equation}
q^4 + (\Delta_h+\kappa^2)q^2 + \Delta_h\kappa^2 - C^2 = 0,
\label{q_poles}
\end{equation}
which determines the dispersion of the dressed environmental modes and encodes the hybridization between the amplitude fluctuation and the screened density-response channel.

For inter-boundary coupling, the relevant regime is the evanescent one, in which propagation along $z$ is exponentially damped. Writing the two decaying branches as
\begin{equation}
q_1=i\gamma_1,\qquad q_2=i\gamma_2,
\qquad \Re\,\gamma_{1,2}>0,
\end{equation}
the inter-plate propagator takes the form
\begin{equation}
G_h^R(\omega,\mathbf{k};a,0)
=\frac{\frac{\kappa^2-\gamma_1^2}{\gamma_1}e^{-\gamma_1 a}-\frac{\kappa^2-\gamma_2^2}{\gamma_2}e^{-\gamma_2 a}}
{2(\gamma_1^2-\gamma_2^2)}.
\label{Gh_interplate}
\end{equation}
This expression shows that inter-boundary correlations are mediated over a finite range set by the inverse decay rates $\gamma_{1,2}$, so that both coherent coupling and correlated noise are exponentially suppressed at large separation.

\subsection{Boundary influence action}\label{boundary_action}

Integrating out the environmental field $h$, which couples linearly to the boundary modes through Eq.~\eqref{S_int_linear}, yields a quadratic induced action for the boundary degrees of freedom. In frequency--momentum space, it can be written as
\begin{equation}
S_{\rm ind}
=-\frac12\int\frac{d\omega\,d^2\mathbf{k}}{(2\pi)^3}\Phi^\dagger(-\omega,-\mathbf{k})\mathbf{\Sigma}^{F}(\omega,\mathbf{k})\Phi(\omega,\mathbf{k}),
\label{S_induced_boundary}
\end{equation}
where
\begin{equation}
\Phi(\omega,\mathbf{k})
=
\begin{pmatrix}
\phi_A(\omega,\mathbf{k})\\[1mm]
\phi_B(\omega,\mathbf{k})
\end{pmatrix},
\end{equation}
and the in-out environmental self-energy matrix is
\begin{equation}
\mathbf{\Sigma}^{F}(\omega,\mathbf{k})
=
\begin{pmatrix}
\lambda_A^2 G_h^{F}(\omega,\mathbf{k};a,a) &
\lambda_A\lambda_B G_h^{F}(\omega,\mathbf{k};a,0)
\\[1mm]
\lambda_A\lambda_B G_h^{F}(\omega,\mathbf{k};0,a) &
\lambda_B^2 G_h^{F}(\omega,\mathbf{k};0,0)
\end{pmatrix}.
\label{Sigma_matrix}
\end{equation}

The diagonal components describe local self-energy corrections, while the off-diagonal components encode the common-environment coupling between the two subsystems. In particular,
\begin{equation}
\Sigma_{AB}^{F}(\omega,\mathbf{k})
=\lambda_A\lambda_B G_h^{F}(\omega,\mathbf{k};a,0)
\label{Sigma_AB}
\end{equation}
controls the inter-boundary physics at the level of the in-out effective action, and its real part generates coherent mediated interactions between the two boundaries.

From the OQS perspective, the causal reduced dynamics is obtained more naturally in the CTP formulation discussed below. In that formulation, the retarded self-energy matrix is obtained from Eq.~\eqref{Sigma_matrix} by replacing $G_h^{F}$ with $G_h^{R}$, while the same environmental spectral data also determine the Hadamard noise kernel. The entire reduced boundary dynamics is therefore controlled by a single environmental Green function evaluated at the boundary positions. Because both diagonal and off-diagonal entries are governed by the same dressed propagator, the two boundary subsystems couple to a common environment and exhibit intrinsically correlated fluctuations. This structure will be important in the following section, where the same spectral support determines the onset of motion-activated excitation production and the associated decohering channel.


\section{Motion-activated excitation-production channel}
\label{imag}

Before constructing the full reduced dynamics, we isolate the environmental excitation channel made available by the relative motion of the two boundary subsystems. In the in-out formulation, it is diagnosed by the imaginary part of the connected vacuum amplitude, which signals the production of real excitations in the common environment. The quantity studied here should be understood as an auxiliary in-out object in which the boundary fluctuations are also contracted. It is therefore of fourth order in the boundary-environment couplings and is not to be identified directly with the CTP noise kernel of the reduced boundary subsystem. Rather, it serves to expose the motion-activated environmental shell that later controls the dominant resonant contribution in the CTP description.

We work in a nonrelativistic surface-response regime $v\ll1$, where the effect of relative motion is incorporated through a Doppler shift of the boundary response function. For the moving plate $A$, this amounts to the substitution
\begin{equation}
G_A^R(\omega,\mathbf{k})\longrightarrow G_A^R(\omega-vk_x,\mathbf{k}),
\label{doppler_GA}
\end{equation}
while the stationary plate $B$ retains
\begin{equation}
G_B^R(\omega,\mathbf{k})
=\frac{1}{-(\omega+i0^+)^2/u_\phi^2 + k_\parallel^2 + m_\phi^2 u_\phi^2}.
\label{GB_ret}
\end{equation}
To make the kinematic origin of the excitation channel more transparent, it is useful to examine the spectral structure of the boundary modes. 

The resonance condition follows directly from the spectral representation. Both boundary propagators take the common form
\begin{equation}
G_\phi^R(\omega,\mathbf{k})
=\frac{1}{-(\omega+i0^+)^2/u_\phi^2 + k_\parallel^2 + m_\phi^2 u_\phi^2},
\end{equation}
with spectral density
\begin{equation}
\begin{aligned}
\rho_\phi(\omega,\mathbf{k})=&-2\,{\rm Im}\,G_\phi^R(\omega,\mathbf{k})\\
=&\frac{\pi u_\phi^2}{\Omega_{\mathbf{k}}}\left[\delta(\omega-\Omega_{\mathbf{k}})-\delta(\omega+\Omega_{\mathbf{k}})\right],
\end{aligned}
\label{rho_phi}
\end{equation}
where $\phi=A,B$. For later convenience, we separate positive- and negative-frequency components,
\begin{equation}
\rho_\phi^{(\pm)}(\omega,\mathbf{k})
=\frac{\pi u_\phi^2}{\Omega_{\mathbf{k}}}\delta(\omega \mp \Omega_{\mathbf{k}}).
\label{rho_pm}
\end{equation}
where the upper (lower) sign corresponds to the positive- (negative-) frequency component.

At leading nontrivial order in the boundary--environment coupling, the dissipative contribution arises from the cut of the common-environment diagram and can be written in spectral form as
\begin{equation}
\begin{aligned}
{\rm Im}\,\Gamma_{\rm cross}
=&\frac{\lambda_A^2\lambda_B^2}{2}
\int\frac{d\omega\,d^2\mathbf{k}}{(2\pi)^3}\,
\rho_A^{(-)}(\omega-vk_x,\mathbf{k})\,
\rho_B^{(+)}(\omega,\mathbf{k}) \\
&\times\big|G_h^R(\omega,\mathbf{k};a,0)\big|^2.
\end{aligned}
\label{ImGamma_spectral_new}
\end{equation}
This expression measures the overlap between the negative-frequency sector of the Doppler-shifted moving boundary and the positive-frequency sector of the stationary boundary, weighted by propagation through the common environment. Since the boundary fluctuations are also contracted, it should be interpreted as a diagnostic of the resonant excitation channel rather than as the reduced-system noise kernel itself.

Before specializing to identical boundary dispersions, the support of the two spectral densities imposes the on-shell conditions
\begin{equation}
\omega = \Omega_B(\mathbf{k}),
\qquad
\omega - vk_x = -\Omega_A(\mathbf{k}),
\end{equation}
and hence the general resonance condition
\begin{equation}
vk_x=\Omega_A(\mathbf{k})+\Omega_B(\mathbf{k}).
\label{general_resonance_condition}
\end{equation}
The corresponding critical velocity is therefore
\begin{equation}
v_c=\inf_{k_x>0,\,k_y}
\frac{\Omega_A(\mathbf{k})+\Omega_B(\mathbf{k})}{k_x}.
\label{general_velocity_threshold}
\end{equation}
For the identical boundary dispersions,
$\Omega_A=\Omega_B=\Omega_{\mathbf{k}}$, Eq.~\eqref{general_resonance_condition} reduces to
\begin{equation}
vk_x=2\Omega_{\mathbf{k}}=2u_\phi \sqrt{k_x^2+k_y^2+m_\phi^2u_\phi^2}.
\label{res_condition_new}
\end{equation}
Solving for $k_x$ gives
\begin{equation}
k_x^\ast=\frac{2u_\phi}{\sqrt{v^2-4u_\phi^2}}\sqrt{k_y^2+m_\phi^2u_\phi^2},
\label{kxstar_new}
\end{equation}
which is real only if
\begin{equation}
v>2u_\phi.
\label{velocity_threshold_new}
\end{equation}
This condition defines the ideal kinematic onset for the identical-dispersion model and follows directly from Eq.~\eqref{boundary_dispersion}. The factor of two is specific to the identical-branch model, which is consistent with the anomalous-Doppler interpretation of quantum \v{C}erenkov radiation~\cite{Maghrebi2013}.

\begin{figure}[t]
\centering
\includegraphics[width=1\linewidth]{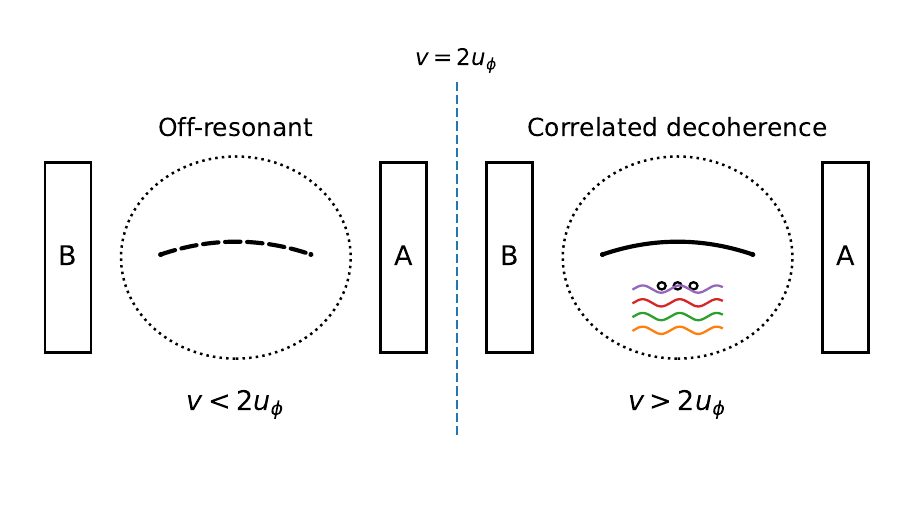}
\caption{Schematic of the velocity-controlled leading resonant contribution to correlated decoherence. The dashed line marks the
ideal identical-dispersion onset at $v=2u_\phi$.}
\label{mechanism}
\end{figure}

\begin{figure}[t]
\centering
\includegraphics[width=1\linewidth]{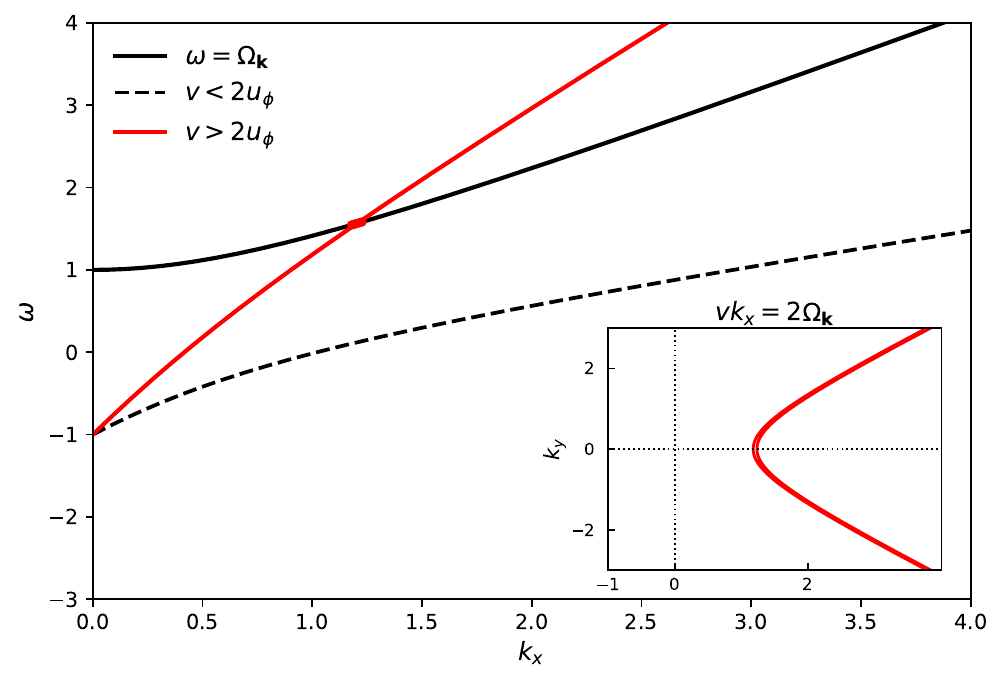}
\caption{Spectral structure of the identical-dispersion model. The branches $\omega=\Omega_{\mathbf{k}}$ and
$\omega=vk_x-\Omega_{\mathbf{k}}$ intersect above the ideal onset $v=2u_\phi$, producing the momentum-space shell shown in
the inset.}
\label{spectral}
\end{figure}

\begin{figure}[t]
\centering
\includegraphics[width=1\columnwidth]{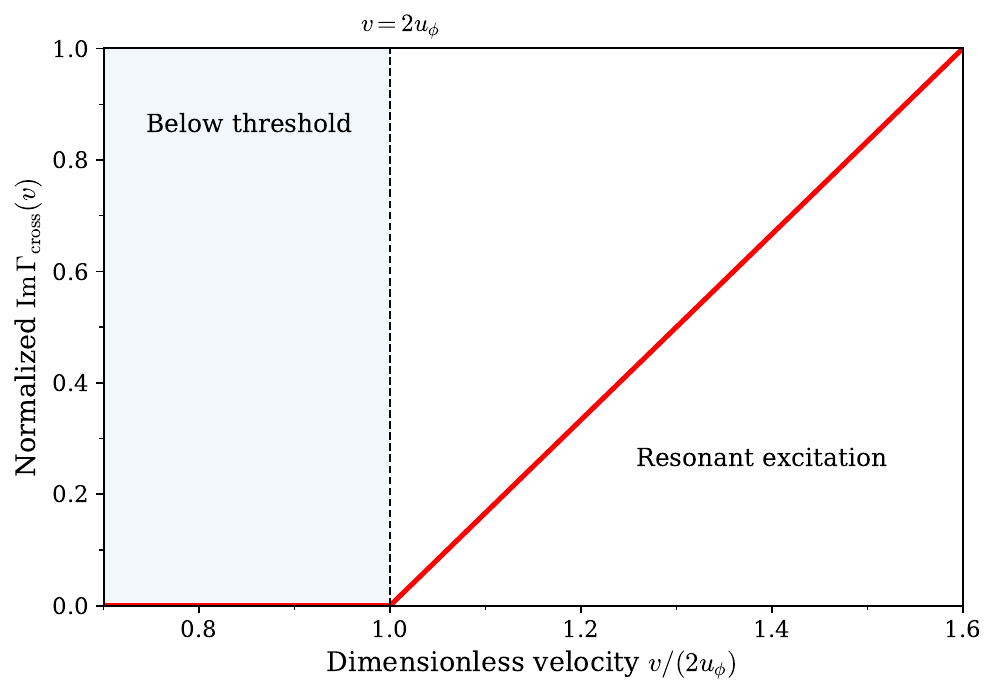}
\caption{Leading resonant contribution to ${\rm Im}\,\Gamma_{\mathrm{cross}}$ as a function of velocity in the identical-dispersion, narrow-linewidth limit. The onset is at $v=2u_\phi$.}
\label{Rv_single}
\end{figure}

Figure~\ref{mechanism} summarizes the velocity-controlled opening of the leading resonant channel, while Figure~\ref{spectral} shows the dispersion $\omega=\Omega_{\mathbf{k}}$ and the Doppler-shifted branch $\omega=vk_x-\Omega_{\mathbf{k}}$. 
In the zero-linewidth limit, the two branches do not intersect for $v<2u_\phi$, whereas a resonant shell appears for $v>2u_\phi$. 
The resulting velocity dependence of the leading resonant contribution to ${\rm Im}\,\Gamma_{\rm cross}$ is shown in
Figure~\ref{Rv_single}.


Using the on-shell constraints, the $\omega$ and $k_x$ integrals can be carried out explicitly, yielding the reduced one-dimensional form
\begin{equation}
{\rm Im}\,\Gamma_{\rm cross}
=\frac{\lambda_A^2\lambda_B^2 u_\phi^2}{8\pi v}
\int_{-\infty}^{\infty}
\frac{dk_y}{k_y^2+m_\phi^2u_\phi^2}
\big|G_h^R(\omega_\ast,\mathbf{k}_\ast;a,0)\big|^2,
\label{ImGamma_reduced_new}
\end{equation}
where
\begin{equation}
\omega_\ast=\frac{u_\phi v}{\sqrt{v^2-4u_\phi^2}}
\sqrt{k_y^2+m_\phi^2u_\phi^2},
\qquad
\mathbf{k}_\ast=(k_x^\ast,k_y).
\label{omega_star_new}
\end{equation}
Details of the resonance reduction and the associated Jacobian are given in Appendix~\ref{reduced_formula}.

Substituting the inter-plate propagator in Eq.~\eqref{Gh_interplate}, we arrive at
\begin{equation}
\begin{aligned}
{\rm Im}\,\Gamma_{\rm cross}
=&\frac{\lambda_A^2\lambda_B^2 u_\phi^2}{8\pi v}
\int_{-\infty}^{\infty}\frac{dk_y}{k_y^2+m_\phi^2u_\phi^2}\,\frac{1}{4(\gamma_1^2-\gamma_2^2)^2}\\
&\times\left|\frac{\kappa^2-\gamma_1^2}{\gamma_1}e^{-\gamma_1 a}-\frac{\kappa^2-\gamma_2^2}{\gamma_2}e^{-\gamma_2 a}\right|^2_{\rm shell}
\end{aligned}
\label{ImGamma_final_new}
\end{equation}
with all quantities evaluated on the resonance shell defined by Eq.~\eqref{res_condition_new}. Thus, in the identical-dispersion and zero-linewidth limit, this particular contribution is nonzero only for $v>2u_\phi$ and remains short-ranged because of the exponential decay inherited from $G_h^R(a,0)$. It is an auxiliary diagnostic of the environmental spectral support rather than the full reduced-system noise kernel. The same resonance condition selects the leading motion-enhanced contribution in the CTP description below.


\section{Reduced boundary dynamics and common-environment kernels}
\label{reduced_dynamics}

We now formulate the boundary dynamics within the framework of OQSs. The relevant object is the reduced density matrix
\begin{equation}
\rho_r(T)={\rm Tr}_{\rm env}\,\rho_{\rm tot}(T),
\label{rho_reduced_def_new}
\end{equation}
obtained after tracing out the cavity environment. The aim of this section is to express the reduced dynamics in terms of the environmental kernels introduced above and to identify the correlated noise channel induced by the common environment. Unlike the in-out quantity ${\rm Im}\,\Gamma_{\rm cross}$ analyzed in Eq.~\eqref{ImGamma_spectral_new}, the CTP formulation directly governs the observable decoherence dynamics of the reduced boundary subsystem.

In frequency--momentum space, the influence action takes the matrix form
\begin{equation}
S_{\rm IF}=-\frac12
\sum_{\alpha,\beta=\pm}
\int \frac{d\omega\,d^2\mathbf{k}}{(2\pi)^3}\,
\Phi_\alpha^\dagger(-\omega,-\mathbf{k})\,
\mathbf{G}_{\alpha\beta}(\omega,\mathbf{k})\,
\Phi_\beta(\omega,\mathbf{k}),
\label{SIF_matrix_form}
\end{equation}
where
\begin{equation}
\Phi_\alpha(\omega,\mathbf{k})
=
\begin{pmatrix}
\lambda_A \phi_A^\alpha(\omega,\mathbf{k})\\[1mm]
\lambda_B \phi_B^\alpha(\omega,\mathbf{k})
\end{pmatrix},
\end{equation}
and $\mathbf{G}_{\alpha\beta}$ denotes the matrix of CTP environmental correlators evaluated at the two interfaces,
\begin{equation}
\mathbf{G}_{\alpha\beta}(\omega,\mathbf{k})
=
\begin{pmatrix}
G_{\alpha\beta}(\omega,\mathbf{k};a,a) &
G_{\alpha\beta}(\omega,\mathbf{k};a,0)
\\[1mm]
G_{\alpha\beta}(\omega,\mathbf{k};0,a) &
G_{\alpha\beta}(\omega,\mathbf{k};0,0)
\end{pmatrix}.
\label{G_ctp_matrix}
\end{equation}
Introducing the average and difference variables
\begin{equation}
\phi_i^\Sigma=\frac{\phi_i^+ + \phi_i^-}{2},
\qquad
\phi_i^\Delta=\phi_i^+ - \phi_i^-,
\qquad
i=A,B,
\label{SigmaDelta_fields}
\end{equation}
and
\begin{equation}
\Phi_X(\omega,\mathbf{k})
=
\begin{pmatrix}
\lambda_A \phi_A^X(\omega,\mathbf{k})\\[1mm]
\lambda_B \phi_B^X(\omega,\mathbf{k})
\end{pmatrix},
\qquad X=\Sigma,\Delta .
\end{equation}
the influence action becomes
\begin{equation}
S_{\rm IF}
=\int \frac{d\omega\,d^2\mathbf{k}}{(2\pi)^3}\left[\Phi_\Delta^\dagger\mathbf{D}_R\Phi_\Sigma+\frac{i}{2}\Phi_\Delta^\dagger\mathbf{N}\,\Phi_\Delta\right],
\label{SIF_DRN_matrix}
\end{equation}
where $\mathbf{D}_R$ and $\mathbf{N}$ are the retarded and noise kernel matrices, respectively. The retarded kernel governs the causal coherent mediation induced by the common environment, whereas the noise kernel encodes the corresponding fluctuations and decohering effects. The noise kernel has the matrix structure
\begin{equation}
\mathbf{N}
=
\begin{pmatrix}
N_{AA} & N_{AB}
\\[1mm]
N_{BA} & N_{BB}
\end{pmatrix},
\label{N_matrix}
\end{equation}
where the diagonal entries describe local environmental noise, while the off-diagonal entry $N_{AB}=N_{BA}^\ast$ encodes correlated noise generated by the common environment. 
\begin{figure}[t]
\centering
\includegraphics[width=1\linewidth]{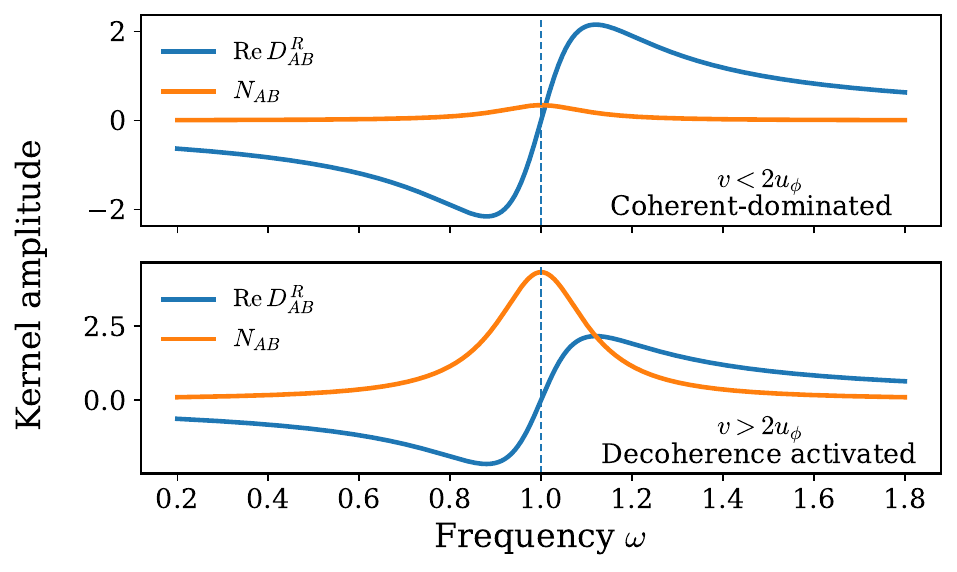}
\caption{Frequency-resolved common-environment kernels below and above the identical-dispersion onset $v=2u_\phi$. The retarded kernel $\mathrm{Re}\,D^R_{AB}$ remains finite across the onset, while the $N_{AB}$ curve shows its boundary-filtered resonant part.}
\label{kernel_structure}
\end{figure}
At zero temperature and in the stationary regime, the cross-noise kernel is determined by the symmetrized (Hadamard) correlator of the dressed environmental mode~\cite{Calzetta2008,Kamenev2011},
\begin{equation}
N_{AB}(\omega,\mathbf{k})= \lambda_A\lambda_B\, G_h^H(\omega,\mathbf{k};a,0),
\end{equation}
with
\begin{equation}
G_h^H(\omega,\mathbf{k};a,0)=\frac12\big(G_h^>(\omega,\mathbf{k};a,0)+G_h^<(\omega,\mathbf{k};a,0)\big).
\end{equation}
Figure~\ref{kernel_structure} illustrates the different velocity dependence of coherent mediation and of the boundary-filtered resonant noise contribution. The bare common-environment kernel in Eq.~\eqref{NAB_Hadamard} contains no explicit velocity and is generally nonzero below $2u_\phi$. Motion enters when it is contracted with the moving-boundary histories or spectral filters. It is this filtered, leading on-shell contribution that is absent below the ideal onset and enhanced when the Doppler-shifted boundary spectra overlap.

Only the full noise matrix $\mathbf{N}$ entering the quadratic influence functional is required to be positive semidefinite, whereas the cross component $N_{AB}$ alone need not be positive definite. Using the spectral representation, one may equivalently write~\cite{Calzetta2008,Kamenev2011}
\begin{equation}
G_h^H(\omega,\mathbf{k};a,0)=\frac12 \coth\!\left(\frac{\beta\omega}{2}\right)\rho_h(\omega,\mathbf{k};a,0),
\label{Hadamard_FDT}
\end{equation}
which at zero temperature reduces to
\begin{equation}
G_h^H(\omega,\mathbf{k};a,0)=\frac12\,{\rm sgn}(\omega)\,\rho_h(\omega,\mathbf{k};a,0).
\label{NAB_Hadamard}
\end{equation}
Here $\rho_h=i(G_h^R-G_h^A)$ is the spectral function of the dressed amplitude mode. We adopt a convention in which $G^>$ and $G^<$ are defined without an explicit prefactor of $-i$, so that $G_h^H$ is real and $\rho_h=i(G_h^R-G_h^A)$. Eq.~\eqref{NAB_Hadamard} therefore gives the exact Gaussian representation of the reduced-system cross-noise kernel in terms of the environmental Hadamard correlator. For boundary histories concentrated near the motion-activated resonant sector, its leading contribution is supported on the same Doppler-shifted shell defined by Eq.~\eqref{res_condition_new}.

For boundary histories with $\Phi_\Delta\neq0$, the imaginary part of the influence action is
\begin{equation}
{\rm Im}\,S_{\rm IF}=\frac12\int \frac{d\omega\,d^2\mathbf{k}}{(2\pi)^3}
\Phi_\Delta^\dagger(\omega,\mathbf{k})
\mathbf{N}(\omega,\mathbf{k})
\Phi_\Delta(\omega,\mathbf{k}),
\label{ImSIF_general_matrix}
\end{equation}
so that off-diagonal elements of the reduced density matrix are suppressed as
\begin{equation}
\rho_r^{\rm off}(T)\sim\exp\!\left[-\mathcal{D}(T)\right]\rho_r^{\rm off}(0),
\label{rho_off_decay_new}
\end{equation}
where $\mathcal{D}(T)\equiv {\rm Im}\,S_{\rm IF}(T)$.

The observable consequence can be expressed more directly in terms of the coherence factor
\begin{equation}
C(T)\equiv \frac{|\rho_r^{\rm off}(T)|}{|\rho_r^{\rm off}(0)|},
\label{coherence_factor_def}
\end{equation}
which in the present Gaussian description behaves as
\begin{equation}
C(T)\simeq e^{-\mathcal{D}(T)}.
\label{coherence_factor_decay}
\end{equation}
An experimentally useful way to isolate the correlated part is to compare common- and differential-mode coherences~\cite{vonLupke2020}. For equal-amplitude histories with $\Delta_A=\pm\Delta_B$, define
\begin{equation}
\mathcal{D}_{\pm}=\mathcal{D}_{AA}+\mathcal{D}_{BB}
\pm2\,\mathrm{Re}\,\mathcal{D}_{AB}.
\label{common_differential_decoherence}
\end{equation}
The local contributions cancel in the difference,
\begin{equation}
\mathcal{D}_{+}-\mathcal{D}_{-}=4\,\mathrm{Re}\,\mathcal{D}_{AB},
\label{differential_cross_extraction}
\end{equation}
so joint Ramsey or echo measurements can extract the cross term without assigning the entire decay to motion. This differential strategy complements the single-probe motion-dependent decoherence observable~\cite{Viotti2021}. The full cross contribution is governed by $N_{AB}$ and is generally finite on both sides of the ideal onset. Its motion-selected resonant part is enhanced when the shell opens.

To make the resonant contribution explicit, we consider boundary histories concentrated near a shell point $(\omega_0,\mathbf{k}_0)$,
\begin{equation}
\phi_i^\Delta(t,\mathbf{x}_\parallel)
=
\Delta_i\,\chi_T(t)\chi_\parallel(\mathbf{x}_\parallel)
e^{-i\omega_0 t+i\mathbf{k}_0\cdot \mathbf{x}_\parallel},
\label{history_windowed_resonant}
\end{equation}
where $i=A,B$. $\chi_T$ and $\chi_\parallel$ are smooth envelope functions and $(\omega_0,\mathbf{k}_0)$ lies in the resonant neighborhood of the motion-activated shell. The cross contribution to the damping functional is then
\begin{equation}
\begin{aligned}
\mathcal{D}_{AB}(T)
=&\int \frac{d\omega\,d^2\mathbf{k}}{(2\pi)^3}
\mathrm{Re}\!\Big[
\lambda_A \phi_A^\Delta(-\omega,-\mathbf{k}) \\
&\times G_h^H(\omega,\mathbf{k};a,0)
\lambda_B \phi_B^\Delta(\omega,\mathbf{k})\Big].
\end{aligned}
\label{gamma_dec_general}
\end{equation}
In the long-time and large-area limit, the envelope functions become sharply peaked around $(\omega_0,\mathbf{k}_0)$ and project onto the corresponding local spectral sector of the environment. For histories concentrated near the motion-activated resonant shell, this support collapses onto the same Doppler-shifted shell defined by Eq.~\eqref{res_condition_new}. The cross damping functional therefore takes the form
\begin{equation}
\mathcal{D}_{AB}(T)\simeq\Delta_A\Delta_B\,T\,\mathcal{A}\,R_{\rm res}^{(AB)},
\label{DAB_def}
\end{equation}
where $\mathcal{A}$ is the effective transverse area selected by the envelope and $R_{\rm res}^{(AB)}$ is the corresponding intensive resonant decoherence-rate density. Equation~\eqref{DAB_def} applies to histories concentrated near the motion-activated shell. More generally, the retarded component of the common-environment kernel controls coherent mediation, while its Hadamard component encodes correlated noise, including off-resonant contributions through $N_{AB}$.

\section{Connection between excitation production and decoherence}\label{decoherence_connection}

We now connect the excitation-production channel identified in the in-out formulation with the decoherence dynamics obtained from the reduced OQS description. The quantities ${\rm Im}\,\Gamma_{\rm cross}$ and the leading resonant contribution to the cross-decoherence exponent are not identical. For stationary histories, we denote the associated decoherence-rate density by $R_{\rm res}^{(AB)}$. The former is defined from the connected vacuum amplitude in the in-out formalism, whereas the latter arises from the noise sector of the CTP influence functional. Within the Gaussian weak-coupling approximation, however, both are controlled by the same environmental spectral information and are dominated by the same motion-activated resonant shell.

Both quantities are governed by the same dressed propagator at the boundaries and therefore probe the same resonant channel, albeit in different observables. As shown by Eqs.~\eqref{ImGamma_spectral_new}, \eqref{res_condition_new}, and \eqref{NAB_Hadamard}, ${\rm Im}\,\Gamma_{\rm cross}$ isolates the on-shell excitation channel of the environment, while the cross-noise kernel $N_{AB}$ determines how the same shell enters the reduced dynamics. For boundary histories concentrated near the free boundary resonances, the leading resonant contribution to the cross-decoherence exponent is supported on the shell defined by Eq.~\eqref{res_condition_new}. In this restricted resonant sense, the corresponding rate density inherits the spectral scaling
\begin{equation}
\begin{aligned}
R_{\rm res}^{(AB)}
\propto
&\lambda_A^2\lambda_B^2
\int\frac{d\omega\,d^2\mathbf{k}}{(2\pi)^3}
\rho_A^{(-)}(\omega-vk_x,\mathbf{k})\\
&\times \rho_B^{(+)}(\omega,\mathbf{k}) \big|G_h^R(\omega,\mathbf{k};a,0)\big|^2,
\end{aligned}
\label{Rres_scaling_spectral}
\end{equation}
up to history-dependent prefactors set by the boundary trajectories. This is not a universal expression for the reduced-system noise kernel itself, but only for its leading resonant contribution.

Using Eq.~\eqref{res_condition_new}, this contribution is supported on the shell $vk_x=2\Omega_{\mathbf{k}}$ and therefore exists only for
\begin{equation}
v>2u_\phi.
\end{equation}
Below threshold, the spectral supports do not overlap and the leading resonant cross-decoherence channel is absent. Above threshold, the resonant shell opens and the common environment becomes an efficient source of correlated decoherence.

Since both ${\rm Im}\,\Gamma_{\rm cross}$ and the leading resonant contribution to $R_{\rm res}^{(AB)}$ are controlled by the same inter-plate propagator $G_h^R(a,0)$, they inherit the same exponential attenuation with separation. They also depend on the condensate parameters through the same interplay of effective coupling and screening: increasing the condensate amplitude enhances the effective boundary coupling $\lambda_{A,B}=g_{A,B}\rho_0$, while also increasing the screening scale $m_A$ and shortening the correlation length. The correct interpretation is therefore not that ${\rm Im}\,\Gamma_{\rm cross}$ equals the decoherence rate, but that it identifies the same on-shell environmental channel controlling the dominant resonant contribution to the noise sector. In this restricted resonant sense, excitation production and decoherence are complementary manifestations of the same underlying spectral channel.


\section{Comparison of condensed and uncondensed environmental phases}
\label{comparison}

It is instructive to compare the condensed phase analyzed above with the uncondensed phase characterized by $\rho_0=0$. From the OQS perspective, the two phases correspond to qualitatively different realizations of a common environment and therefore modify the effective kernels governing the reduced boundary dynamics in different ways.

In the uncondensed phase, the boundary fields couple to the composite density operator of the bulk field,
\begin{equation}
S_{\rm int}^{\rm uncond}\sim g_A \phi_A |\psi|^2\big|_{z=a}+g_B \phi_B |\psi|^2\big|_{z=0}.
\label{Sint_uncondensed}
\end{equation}
The system-environment interface therefore probes a quadratic bulk operator, so that the influence functional is governed by composite two-particle correlators rather than by a single elementary propagator, even in a Gaussian bulk state.

In contrast, in the condensed phase the density operator admits the expansion
\begin{equation}
|\psi|^2=\frac{\rho_0^2}{2}+\rho_0 h+\frac12 h^2,
\end{equation}
so that, within the Gaussian approximation, the leading system--environment coupling becomes linear in the amplitude fluctuation,
\begin{equation}
S_{\rm int}^{\rm cond}=\int dt d^2\mathbf{x}_\parallel\left[\lambda_A\phi_A h|_{z=a}+\lambda_B\phi_B h|_{z=0}\right].
\label{Sint_condensed_comp}
\end{equation}
The environmental influence is therefore controlled by a single dressed mode $h$, with both the retarded kernel $\mathbf{D}_R$ and the noise kernel $\mathbf{N}$ constructed from its Green functions evaluated at the boundary positions.

The difference between the two phases is also reflected in the spatial propagation properties of the environment. In the uncondensed phase, inter-plate propagation is governed primarily by the bare bulk gap scale,
\begin{equation}
G_{\rm uncond}(z,z')\sim\frac{e^{-\tilde\kappa|z-z'|}}{2\tilde\kappa},
\label{Guncond_generic}
\end{equation}
where
\begin{equation}
\tilde\kappa=\sqrt{k_\parallel^2+M^2-\omega^2/u_\psi^2},
\end{equation}
so that the correlation length is set directly by the scalar mass $M$. In the condensed phase, by contrast, propagation is governed by the dressed propagator $G_h^R(a,0)$, whose attenuation scale depends on both the amplitude gap and the screening scale $m_A$, leading to a shorter-ranged but more structured correlation profile. Despite these differences, the kinematic threshold remains $v>2u_\phi$, since it is determined by the overlap of boundary spectral densities rather than by the detailed structure of the environment. The environmental phase therefore does not alter the condition for opening the motion-induced channel, but only its strength, spectral weight, and spatial range once activated.

Within the present effective description, the two phases correspond to distinct types of common environments. In the uncondensed phase, the boundary subsystem couples to a composite bath with less sharply resolved spectral features. In the condensed phase, a dominant dressed mode leads to a more transparent kernel structure, while screening reduces the correlation length and suppresses long-range transmission of both interaction and decoherence. The velocity threshold, however, remains controlled by boundary kinematics and is not altered by the environmental phase. The environmental phase controls the efficiency of the activated channel, but not its kinematic onset. A complementary Gaussian benchmark in the symmetric phase is summarized in Appendix~\ref{symmetric_phase}.


\section{Decoherence rate and parameter dependence}\label{decoherence_rate}

\subsection{Explicit decoherence rate}
\label{explicit_rate}

For numerical evaluation, we replace the ideal boundary spectral densities by Lorentzian-broadened forms with a small width $\Gamma_\phi$,
\begin{equation}
\rho_\phi^{(\pm)}(\omega,\mathbf{k})\longrightarrow
\rho_{\phi,\Gamma}^{(\pm)}(\omega,\mathbf{k}),
\end{equation}
where the broadened spectral functions are centered at $\omega=\pm\Omega_{\mathbf{k}}$ and reduce to the delta-function expressions in Eq.~\eqref{rho_pm} as $\Gamma_\phi\to0$. This smooths the ideal resonance shell and yields a finite rate density.

The corresponding regularized leading contribution to the cross-decoherence rate may then be written schematically as
\begin{equation}
\begin{aligned}
R_{\Gamma}^{(AB)}
=&\lambda_A^2\lambda_B^2
\int\frac{d\omega\,d^2\mathbf{k}}{(2\pi)^3}
\rho_{A,\Gamma}^{(-)}(\omega-vk_x,\mathbf{k})
\rho_{B,\Gamma}^{(+)}(\omega,\mathbf{k})\\
&\times\big|G_h^R(\omega,\mathbf{k};a,0)\big|^2.
\end{aligned}
\label{Rgamma_regularized}
\end{equation}

For unequal boundary linewidths, the convolution is controlled by the total overlap width
\begin{equation}
\Gamma_\Sigma=\Gamma_A+\Gamma_B
\label{total_overlap_width}
\end{equation}
and by the detuning
\begin{equation}
\Delta(\mathbf{k};v)=vk_x-\Omega_A(\mathbf{k})-\Omega_B(\mathbf{k}).
\label{resonance_detuning}
\end{equation}
Near a representative shell point $\mathbf{k}_\ast$, the velocity interval over which the ideal onset is rounded scales as
\begin{equation}
\delta v\sim\frac{\Gamma_\Sigma}{|k_x^\ast|}.
\label{velocity_crossover_width}
\end{equation}
Thus any nonzero linewidth produces subthreshold spectral tails. The strict threshold belongs to the zero-linewidth theory, whereas a measurement observes a crossover centered on the same kinematic condition.

For fixed $v$, $a$, and condensate parameters, it is convenient to perform the $\omega$ integral numerically, leaving
\begin{equation}
R_{\Gamma}^{(AB)}(v,a)
=\lambda_A^2\lambda_B^2 \int\frac{d^2\mathbf{k}}{(2\pi)^2} \mathcal{W}_\Gamma(v,\mathbf{k})
\big|G_h^R(\omega,\mathbf{k};a,0)\big|^2,
\label{plot_form}
\end{equation}
where $\mathcal{W}_\Gamma(v,\mathbf{k})$ is the effective broadened overlap weight obtained after the frequency integration. This form makes explicit that the velocity dependence enters through spectral overlap, while the separation dependence is controlled by the inter-plate propagator. 

\begin{figure}[t]
\centering
\includegraphics[width=0.9\columnwidth]{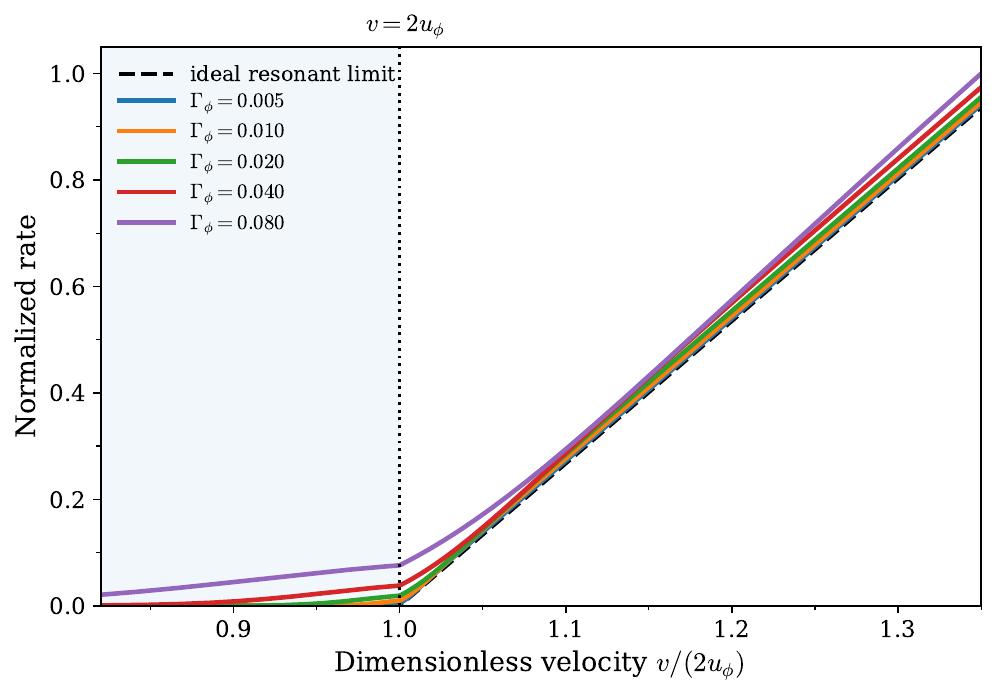}
\caption{Velocity dependence of the regularized resonant cross-decoherence contribution $R_{\Gamma}^{(AB)}(v)$ for different linewidths, together with the zero-linewidth result. }
\label{Gamma_scan}
\end{figure}

Figure~\ref{Gamma_scan} shows the regularized resonant contribution for several linewidths together with the ideal result. Consistent with the identical-boundary model, the numerical curves use $\Gamma_A=\Gamma_B=\Gamma_\phi$, so that $\Gamma_\Sigma=2\Gamma_\phi$. Finite $\Gamma_\Sigma$ generates off-shell subthreshold tails and rounds the zero-width nonanalytic onset. The center of this crossover continues to be set by the kinematic resonance condition, but no strict experimental threshold remains once the spectral functions have finite width.

In the zero-linewidth limit, Eq.~\eqref{Rgamma_regularized} reduces to the shell expressions obtained above. The plots in Sec.~\ref{parameter_dependence} intentionally display this limiting resonant contribution in order to isolate its kinematic origin. Quantitative modeling under experimental conditions instead incorporates the measured linewidths and the crossover scale in Eq.~\eqref{velocity_crossover_width}.


\subsection{Dependence on system parameters}
\label{parameter_dependence}

We now summarize how the leading cross-decohering channel depends on the main control parameters of the system. Based on the scaling form \eqref{Rres_scaling_spectral}, the discussion refers to the dominant resonant contribution within the Gaussian quasi-static approximation. The relevant parameters govern the activation of the resonant channel, the strength of the system--environment coupling, and the spatial range of environmental correlations. The numerical results shown below are obtained by direct evaluation of Eq.~\eqref{plot_form}. Unless otherwise stated, we use the narrow-width limit to isolate the leading resonant contribution.

The relative velocity $v$ controls both the kinematic activation of the decohering channel and the spectral region probed by the reduced dynamics. Through the threshold condition $v>2u_\phi$, it determines whether the resonant shell exists. Below threshold, the spectral supports of the two boundary subsystems do not overlap in the manner required for the motion-activated shell, and the leading resonant cross-decohering contribution is absent. The full noise kernel, however, remains nonzero in general because of off-resonant environmental fluctuations, but without the enhanced on-shell contribution identified in Eq.~\eqref{ImGamma_spectral_new}. Above threshold, the resonance condition is satisfied and the environment can support real excitation processes, thereby activating the dominant decohering channel. Once the channel is open, the velocity also shifts the on-shell frequency and momentum according to Eqs.~\eqref{kxstar_new} and \eqref{omega_star_new}. Varying $v$ therefore selects which region of the environmental spectrum is probed and modulates both the magnitude and the spectral content of the decoherence rate.

\begin{figure}[t]
\centering
\includegraphics[width=1\linewidth]{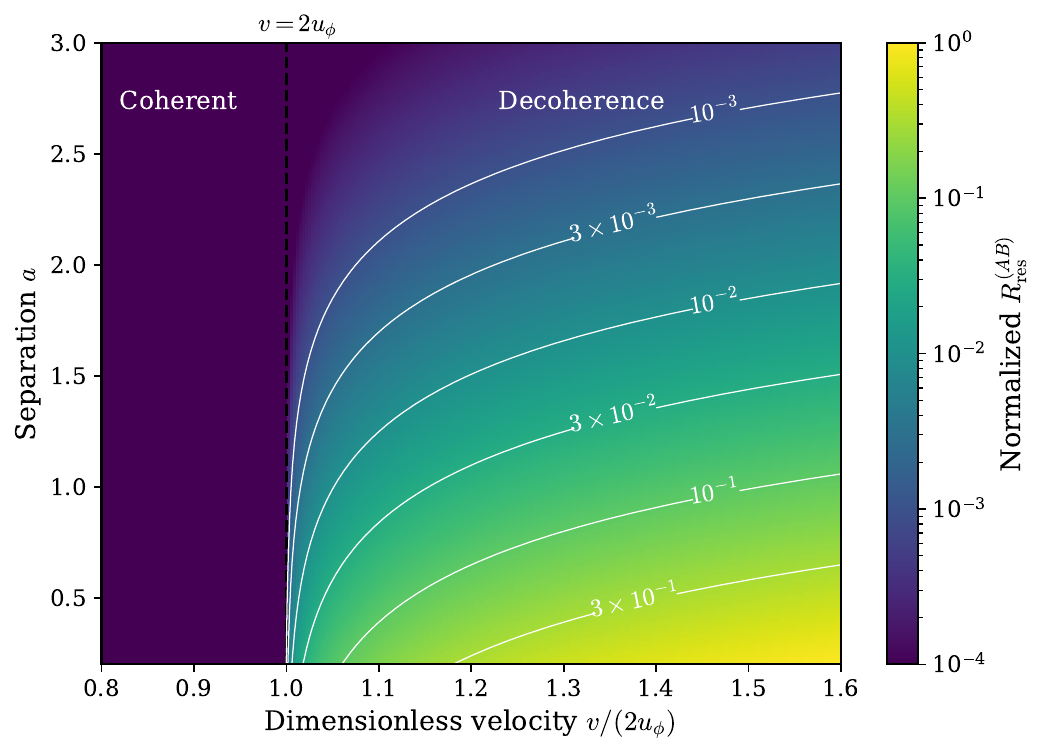}
\caption{Density plot of the leading resonant cross-decoherence rate $R_{\mathrm{res}}^{(AB)}(v,a)$ as a function of
$v/(2u_\phi)$ and separation $a$.}
\label{va_heatmap}
\end{figure}
Figure~\ref{va_heatmap} shows the leading resonant cross-decoherence rate as a function of the dimensionless velocity $v/(2u_\phi)$ and the plate separation $a$. The threshold at $v=2u_\phi$ is independent of separation, confirming its kinematic origin, while above threshold the rate decreases rapidly with increasing separation because of the finite spatial correlation length of the structured environment. The contour pattern further indicates that larger velocity partially compensates for spatial attenuation.

\begin{figure}[t]
\centering
\includegraphics[width=1\linewidth]{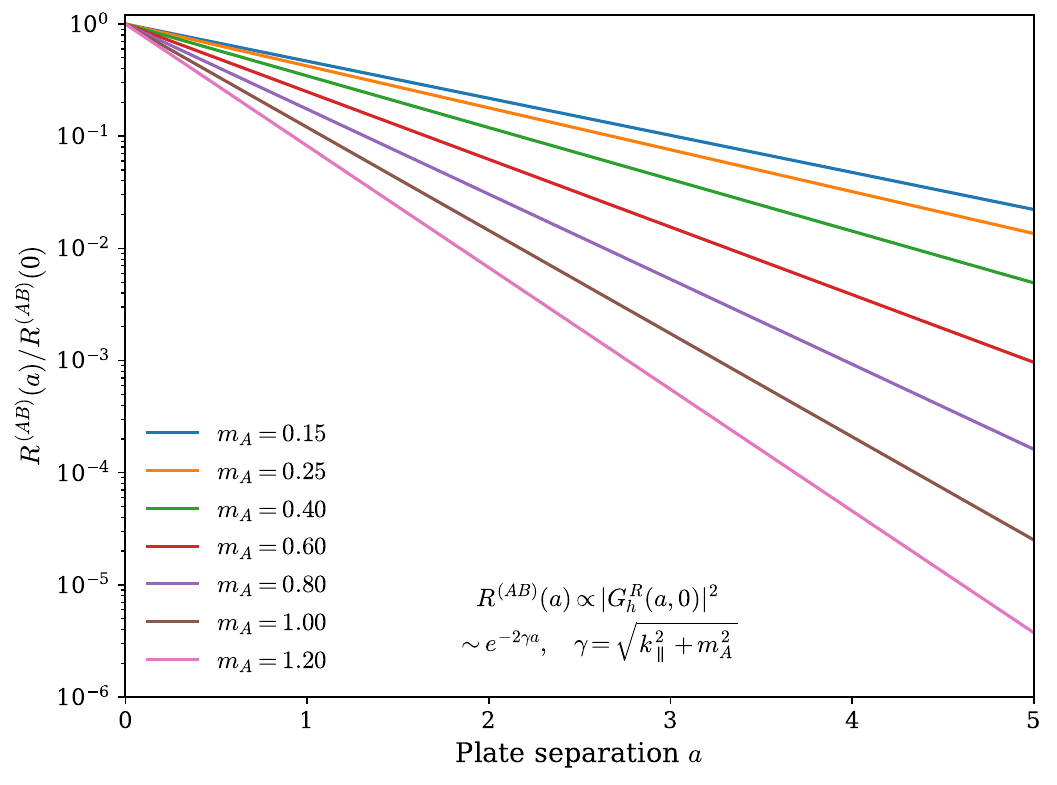}
\caption{Normalized cross-decoherence rate as a function of separation $a$ for different screening scales $m_A$.}
\label{Ra_vs_a}
\end{figure}

We now isolate the dependence on the plate separation $a$. The separation enters through the inter-plate propagator $G_h^R(a,0)$ and therefore controls the spatial range of correlated noise. In the evanescent regime of interest, the leading resonant contribution is exponentially suppressed,
\begin{equation}
R_{\rm res}^{(AB)}\sim e^{-2\,{\rm Re}\,\gamma_{\rm min}\,a},
\end{equation}
where ${\rm Re}\,\gamma_{\rm min}$ denotes the smallest attenuation rate among the dressed modes. As shown in Figure~\ref{Ra_vs_a}, the cross-decoherence rate decays exponentially with separation, confirming that correlated environmental fluctuations are short-ranged. Increasing the screening scale $m_A$ enhances the attenuation and reduces the effective correlation length of the environment. At large separation, the common environment effectively factorizes into two weakly correlated local baths.

\begin{figure}[t]
\centering
\includegraphics[width=1\linewidth]{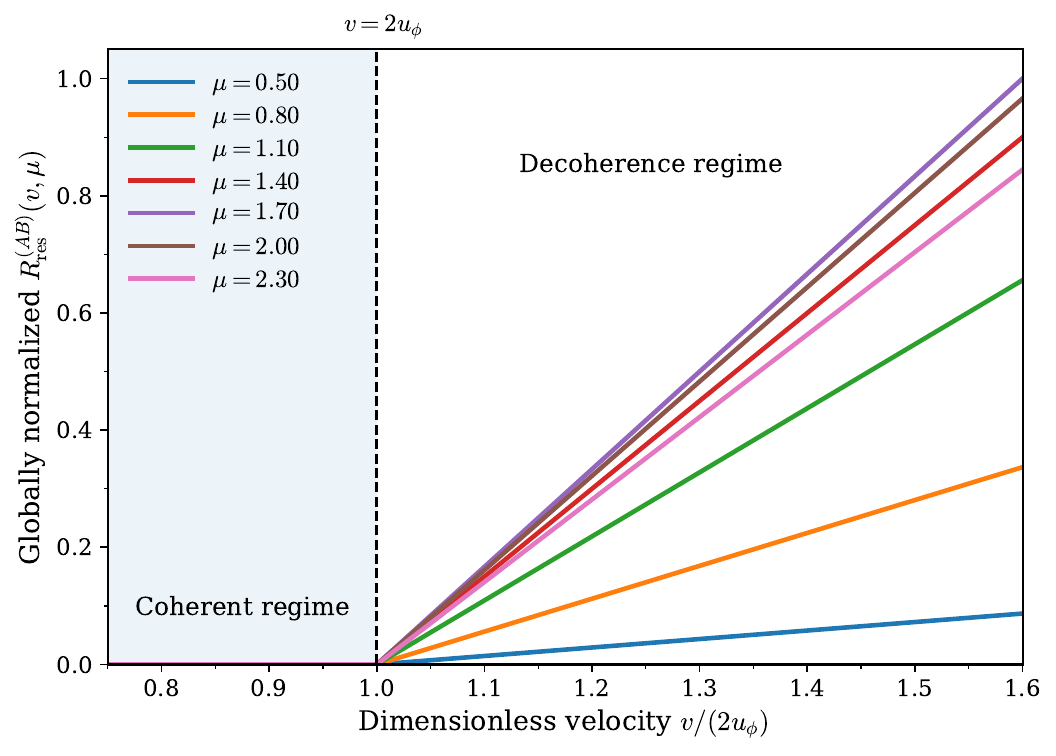}
\caption{
Velocity dependence of the resonant cross-decoherence rate
$R_{\mathrm{res}}^{(AB)}(v,\mu)$ for several values of the chemical potential $\mu$.}
\label{mu_scan}
\end{figure}

\begin{figure}[t]
\centering
\includegraphics[width=1\columnwidth]{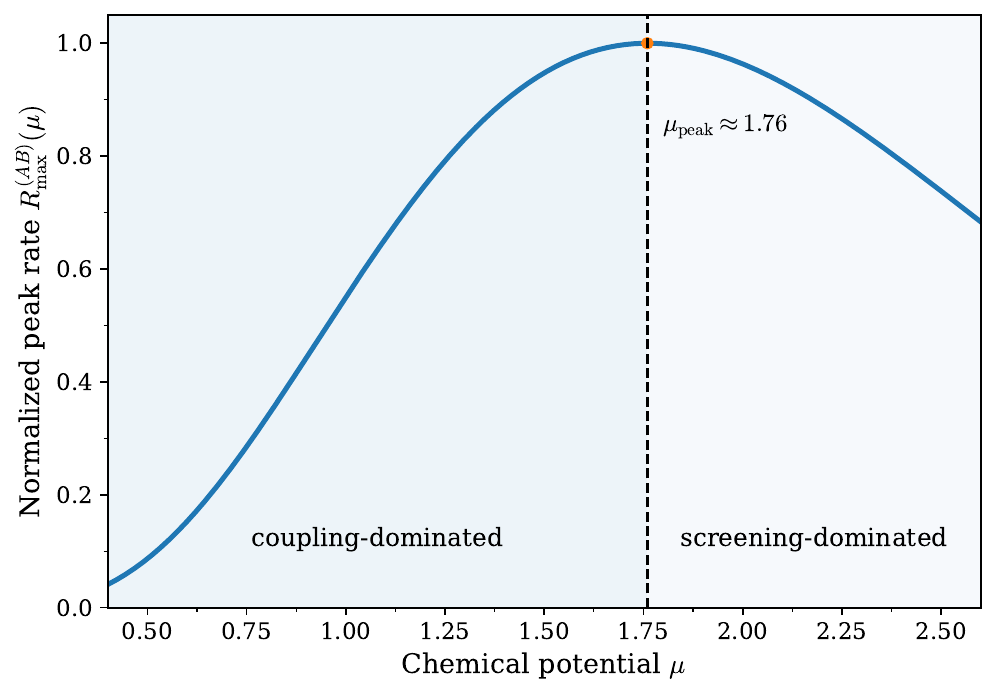}
\caption{Peak value of the resonant cross-decoherence rate $R_{\max}^{(AB)}(\mu)$ as a function of chemical potential $\mu$.}
\label{Rmax_mu}
\end{figure}

The chemical potential $\mu$ affects both the effective coupling strength and the screening properties of the environment through the condensate amplitude,
\begin{equation}
\rho_0^2=\frac{4}{\lambda_\psi}\left(\frac{\mu^2}{u_\psi^2} - m_\psi^2u_\psi^2\right),
\end{equation}
and hence through
\begin{equation}
\lambda_A=g_A\rho_0,
\qquad
\lambda_B=g_B\rho_0,
\end{equation}
and
\begin{equation}
m_A^2=\frac{e^2\rho_0^2}{u_\psi^2}.
\end{equation}
Increasing $\mu$ enhances the system--environment coupling while also increasing the screening strength and reducing the correlation length. The resulting dependence of $R_{\rm res}^{(AB)}$ on $\mu$ is therefore generically non-monotonic. Figure~\ref{mu_scan} shows the velocity dependence of the resonant cross-decoherence rate for several values of $\mu$. The threshold at $v=2u_\phi$ remains unchanged, confirming its kinematic origin, while the above-threshold amplitude first increases with $\mu$ because of stronger coupling and then decreases at larger $\mu$ as screening suppresses inter-boundary correlations. Figure~\ref{Rmax_mu} makes this non-monotonic behavior more explicit by showing the peak value of the resonant cross-decoherence rate as a function of $\mu$. 

These dependencies show that the leading cross-decohering channel is jointly controlled by boundary kinematics and environmental correlation properties. The relative velocity governs the activation of the resonant channel, while the condensate parameters determine its strength and range.


\section{Experimental relevance and possible implementations}
\label{experimental_relevance}

Although the present analysis is formulated within an effective
field-theoretic OQS description, its essential ingredients admit
possible experimental realizations. The mechanism requires a common
structured bosonic environment, a finite propagation length, and a
Doppler-shifted spectral overlap that activates a resonant channel. Among the
available platforms, superconducting and phononic architectures provide a
particularly natural route because synthetic motion can be implemented
without literal mechanical displacement.

Motion-induced effects are already emulated in superconducting circuits
through time-dependent boundary conditions. In particular, the dynamical
Casimir effect has been observed in microwave circuits by modulating the
effective electrical length of a coplanar waveguide with a SQUID termination
\cite{Wilson2011,Lahteenmaki2013}. Structured bosonic environments with finite
propagation length are also available in circuit quantum acoustodynamics and
related phononic platforms, where long-lived acoustic modes with tunable
spectral properties can couple strongly to superconducting qubits
\cite{Chu2017,Manenti2017,Gustafsson2014,OConnell2010,Satzinger2018,Kitzman2023}.
These ingredients make superconducting--phononic devices a promising setting
for the present effect. Propagating microwave and phononic quantum
channels have been realized~\cite{Kurpiers2018,Bienfait2019}, together with
two-qubit correlated-noise spectroscopy~\cite{vonLupke2020}. The complete
protocol remains to be implemented.

A proof-of-principle realization would combine two spatially
separated multimode boundary bands or short resonator arrays with a common
propagating phononic channel~\cite{Bienfait2019}.
Auxiliary superconducting qubits would probe the collective boundary
modes rather than represent the continuum fields $\phi_A$ and
$\phi_B$~\cite{Manenti2017,Kitzman2023}.
In this setting, the
relative velocity is replaced by a synthetic drift velocity generated through
traveling-wave modulation or a moving-frame protocol. 
A traveling modulation with angular frequency $\Omega_m$ and wave number $q_m$ defines the synthetic velocity
\begin{equation}
v_{\rm syn}=\frac{\Omega_m}{q_m}.
\label{synthetic_doppler_velocity}
\end{equation}
This allows the synthetic velocity to be tuned across the ideal
zero-linewidth onset $v_{\rm syn}=2u_\phi$ of the identical-dispersion model.
The correspondence between the effective-theory variables and the
proposed experimental controls and readouts is summarized in
Table~\ref{platform_mapping}.

\begin{table*}[t]
\caption{Effective-theory mapping for the proposed synthetic-motion
superconducting--phononic implementation. }
\begin{ruledtabular}
\begin{tabular}{p{1.4cm}p{8.5cm}p{5cm}}
Quantity & Experimental meaning & Control or readout \\
\hline

$\phi_A,\phi_B$ &
Collective low-energy modes of two spatially separated multimode
boundary bands or short resonator arrays
\cite{Manenti2017,Kitzman2023} &
Auxiliary-qubit band spectroscopy and joint Ramsey or echo readout \\

$h$ &
Propagating modes of the shared phononic waveguide or acoustic network
\cite{Gustafsson2014,Bienfait2019} &
Channel spectroscopy and attenuation measurement \\

$v_{\rm syn}$ &
Phase velocity $\Omega_m/q_m$ of the proposed traveling-wave modulation
that produces the synthetic Doppler shift&
Modulation frequency and wave number \\

$u_\phi$ &
Group or phase velocity of the relevant approximately linear boundary
band &
Extracted from calibrated mode dispersion\\

$a$ &
Separation between the two sampled boundary regions
 &
Device geometry or time-of-flight calibration \cite{Bienfait2019} \\

$R_{\rm res}^{(AB)}$ &
Motion-activated resonant contribution to the correlated dephasing rate &
Velocity-dependent common- and differential-mode coherences or a
spectrally resolved two-node cross-spectrum~\cite{vonLupke2020} \\

\end{tabular}
\end{ruledtabular}
\label{platform_mapping}
\end{table*}

For this architecture, a representative node-separation target is
$a\simeq2\,\mathrm{mm}$~\cite{Bienfait2019}. The representative acoustic velocity is
$u_\phi\simeq4\times10^3\,\mathrm{m\,s^{-1}}$, and the illustrative
modulation wavelength is $2\pi/q_m=0.1$ to $1\,\mathrm{mm}$. The
identical-dispersion onset $v_{\rm syn}=2u_\phi$ then gives
\begin{equation}
\frac{\Omega_{m,c}}{2\pi}
=\frac{2u_\phi}{2\pi/q_m}
\simeq8\text{--}80\,\mathrm{MHz}.
\end{equation}
The wavelength interval and corresponding onset-frequency range are
design targets for the proposed protocol.

The experimentally measurable couplings are the linearized parameters
$\lambda_i=g_i\rho_0$ defined following Eq.~\eqref{S_int_linear}.
For a symmetric proof-of-principle implementation, a representative target is $\lambda_A/(2\pi)\simeq\lambda_B/(2\pi) \simeq5\text{--}12\,\mathrm{MHz}$, motivated by measured qubit--acoustic coupling strengths~\cite{Manenti2017,Kitzman2023}.
These coupling measurements do not determine the microscopic coefficients
$g_A$ and $g_B$. Their values depend on $\rho_0$ and the normalization of the
continuum fields. In an asymmetric implementation, $\lambda_A$ and
$\lambda_B$ are calibrated separately through spectroscopy.

In the proposed readout, $R_{\rm res}^{(AB)}$ corresponds to the motion-selected resonant contribution to the correlated dephasing-rate density. 
The difference between common- and differential-mode Ramsey or echo coherences extracts the full cross contribution, as shown in
Eq.~\eqref{differential_cross_extraction}. 
A calibrated two-node noise cross-spectrum provides an alternative readout of the correlated contribution associated with $N_{AB}$~\cite{vonLupke2020}. 
The resonant part is identified from its dependence on $v_{\rm syn}$ or by selecting the corresponding spectral window. 
For the cross-spectrum readout, the signal-to-noise ratio (SNR) quantifies the visibility of the motion-selected cross-noise enhancement
associated with $R_{\rm res}^{(AB)}$ relative to the local readout background.


The finite-width velocity crossover in
Eq.~\eqref{velocity_crossover_width} maps to the modulation-frequency width
\begin{equation}
\delta f_m\sim
\frac{q_m\Gamma_\Sigma}{2\pi|k_x^\ast|}.
\label{modulation_frequency_width}
\end{equation}
For momentum-matched modulation $q_m\simeq|k_x^\ast|$, this reduces to
$\delta f_m\sim\Gamma_\Sigma/(2\pi)$. Resolving the crossover with at least
five frequency intervals gives the scan-step criterion
\begin{equation}
\Delta f_m
\lesssim\frac{\delta f_m}{5}
\sim\frac{\Gamma_\Sigma}{10\pi}.
\label{modulation_frequency_step}
\end{equation}
The corresponding source-stability target lies below $\Delta f_m$, and
both scales are set by the device linewidths. Increasing $\Gamma_\phi$
broadens the crossover and enhances the subthreshold tail. The smoother onset
reduces the precision of locating its center.

Beyond this concrete implementation, it is useful to compare
characteristic velocity and length scales across other platforms.
Representative parameter regimes across these systems are summarized in
Table~\ref{platform_parameters}. For Bose--Einstein condensates and cold-atom
systems, characteristic velocity scales lie in the mm/s range, while the
relevant obstacle and healing-length scales are of order microns to
submicrons; the values listed in the table are adopted from
Refs.~\cite{Leboeuf2001,Wilson2022,Desbuquois2012}. In graphene and plasmonic
systems, representative propagation velocities span the
$10^{6}$--$10^{7}\ \mathrm{m/s}$ range, while confinement and screening
lengths are typically on the scale of tens of nanometers; the corresponding
table entries are based on
Refs.~\cite{In2022,Goldflam2015,Carbotte2012,Anglhuber2025,Wagner2014}.
These values enable a comparison of characteristic scales across
physically distinct systems.

\begin{table}[t]
\caption{Representative parameter values used for an
illustrative cross-platform comparison. Here $a$ represents a
representative separation, obstacle, or gap scale, while $\xi$ characterizes a
confinement, screening, or propagation length. }
\begin{ruledtabular}
\begin{tabular}{lccc}
Platform & $u_\phi$ (m/s) & $a$ & $\xi$ \\
\hline
BEC~\cite{Leboeuf2001,Wilson2022}
& $2.0\times10^{-3}$ & $4~\mu\mathrm{m}$ & $0.25~\mu\mathrm{m}$ \\
Cold atoms~\cite{Desbuquois2012}
& $1.6\times10^{-3}$ & $2.0~\mu\mathrm{m}$ & $0.3~\mu\mathrm{m}$ \\
Graphene (Doppler)~\cite{In2022,Goldflam2015,Carbotte2012}
& $1.0\times10^{6}$ & $20~\mathrm{nm}$ & $30~\mathrm{nm}$ \\
Plasmonic confinement~\cite{Anglhuber2025,Wagner2014}
& $2.41\times10^{7}$ & $40~\mathrm{nm}$ & $20~\mathrm{nm}$ \\
\end{tabular}
\end{ruledtabular}
\label{platform_parameters}
\end{table}

\begin{figure}[t]
\centering
\includegraphics[width=1.0\linewidth]{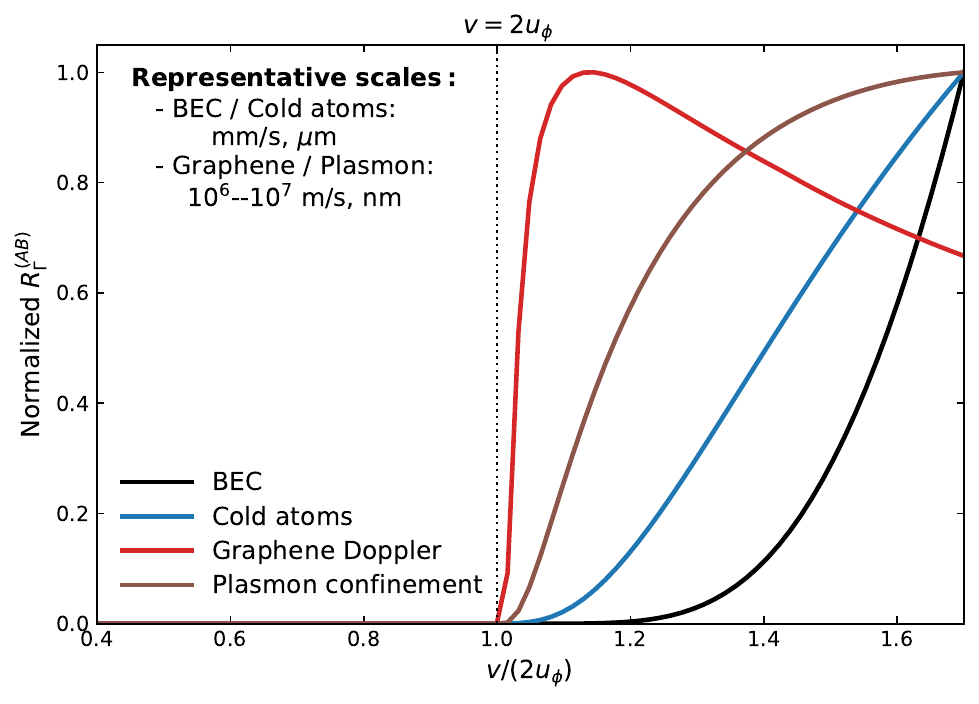}
\caption{Normalized resonant cross-decoherence contribution as a function of
$v/(2u_\phi)$ for representative platform scales, evaluated within the
identical-dispersion model}.
\label{platform_comparison}
\end{figure}

Using these representative parameters, Figure~\ref{platform_comparison}
shows the normalized resonant contribution of the motion-activated
decoherence channel within the identical-dispersion model. The common
alignment of the curves at $v/(2u_\phi)=1$ follows from the chosen model and
normalization. For general dispersions, the onset is determined by the
spectral-overlap criterion in
Eq.~\eqref{general_velocity_threshold}. In synthetic implementations, $v$
denotes the phase velocity of the modulation that generates the effective
Doppler shift. The onset position and above-onset line shape depend on the
platform-specific dispersion, linewidth, and spectral weight.

For ultracold atomic systems, experimental observations of critical
velocities provide a qualitative analogy. Dissipation and heating remain weak below a
system-dependent threshold but rise once phonon or vortex channels become
accessible \cite{Raman1999,Onofrio2000,Engels2007,Desbuquois2012}.
This motion-induced activation is associated with Landau- or
nucleation-type thresholds, and motivates the ultracold-atom comparison in
Fig.~\ref{platform_comparison}.

In graphene and plasmonic systems, a direct measurement of a
velocity-dependent decoherence rate is still lacking, but several closely
related effects have already been identified. Theory predicts drift- and
Doppler-induced modifications of plasmon propagation in current-carrying
graphene and related electron systems \cite{Morgado2018,VanDuppen2016}, while
experimentally relevant propagation velocities and nanoscale confinement
regimes are documented in
Refs.~\cite{In2022,Goldflam2015,Carbotte2012,Anglhuber2025,Wagner2014}.
Near-field optical experiments have also revealed strongly confined plasmon
modes with finite propagation length and pronounced spatial attenuation
\cite{Fei2012,Chen2012,Basov2016,Alonso2017}.
These results identify drift-sensitive propagation, strong confinement,
and finite propagation length as ingredients relevant to the motion-activated
channel studied in the present work.

\section{Conclusion}
\label{conclusion}

We have investigated the reduced dynamics of two spatially separated boundary
subsystems coupled to a common structured environment under relative motion.
Within a Gaussian OQS framework, integrating out the environment yields an
influence functional in which coherent mediation and correlated fluctuations
are governed by the same dressed environmental correlator evaluated at the
boundary positions. This provides a unified description of common-bath
interaction and decoherence in a setting where motion modifies the spectral
matching conditions of the subsystem responses.

We identify the ideal kinematic onset of the leading resonant
contribution to motion-activated correlated decoherence. Relative motion
Doppler-shifts the boundary spectra and opens a resonant shell
when the general spectral-overlap condition is satisfied.
The critical velocity depends on the boundary dispersions. For identical
boundary dispersions, the onset condition becomes $v>2u_\phi$.
Below this onset, the leading resonant contribution to the off-diagonal
decoherence functional is absent in the zero-linewidth limit, although the
full cross-noise kernel generally remains finite. Above it, the resonant shell
enhances the motion-selected cross-noise contribution and the associated
correlated decoherence.
Finite linewidth generates subthreshold tails and rounds the sharp
theoretical onset into a crossover. The resulting enhancement is
therefore governed by the opening of an environmental channel controlled by
spectral overlap rather than by a smooth increase of dissipation alone.

The analysis also clarifies the relation between excitation production in the
in-out formulation and decoherence in the reduced CTP description. Although
these quantities are not identical, within the Gaussian weak-coupling
resonant regime they are controlled by the same on-shell environmental
structure. In the reduced dynamics, the retarded kernel governs coherent
coupling, while the Hadamard kernel determines the correlated noise entering
the decoherence functional. This yields a direct reduced-dynamics signature
of motion-activated correlated decoherence through the suppression of
off-diagonal coherence and the resonant rate density
$R_{\rm res}^{(AB)}$.

The mechanism relies on a common structured bosonic environment, a
finite propagation length, and Doppler-shifted spectral matching, and
is therefore potentially relevant to several experimental platforms.
In particular, the proposed synthetic-motion superconducting--phononic implementation provides a proof-of-principle route. 
The synthetic velocity is experimentally tunable. The quantity $R_{\rm res}^{(AB)}$ corresponds to a
modulation-dependent correlated-dephasing signal. This signal can be measured through joint Ramsey or echo decay or a
two-node noise cross-spectrum. The required ingredients have been demonstrated separately, while the
complete synthetic-motion protocol remains to be implemented.
Within the controlled regime considered, the detailed magnitude of the decoherence rate depends on environmental dressing and resonance broadening.
The spectral-overlap criterion extends beyond the identical-dispersion example. 
The distinction between coherent mediation and correlated decoherence is a general feature of the OQS formulation.
These results show that relative motion, or its synthetic analogue, can be used to activate and enhance the resonant contribution to correlated decoherence in a common environment.


\section*{Acknowledgement}
We gratefully acknowledge fruitful discussions with Zhencheng Fu and Hongwei Tan. Yang Wang was supported by Dongying Science Development Fund(Grant No.DJB2023015). Zhilei Sun was supported by Dongying Science and Technology Development Guidance Program (Grant No.2025ZDJH60). Feiyi Liu, Min Guo and Mingyang Liu were supported by National Natural Science Foundation of China (Grant No.12564032), Yunnan Provincial Department of Education Science Research Fund Project (Grant No.2025J0942, 2026J0977), and Chuxiong Normal College Doctoral Research Initiation Fund Project (Grant No.BSQD2407, BSQD2507).

\bibliography{bibtex}

@article{Farias2016,
  author    = {M. Bel{\'e}n Far{\'\i}as and Fernando C. Lombardo},
  title     = {Dissipation and decoherence effects on a moving particle in front of a dielectric plate},
  journal   = {Physical Review D},
  volume    = {93},
  pages     = {065035},
  year      = {2016},
  month     = mar,
  publisher = {American Physical Society},
  doi       = {10.1103/PhysRevD.93.065035},
  url       = {https://doi.org/10.1103/PhysRevD.93.065035}
}

@article{Viotti2019,
  author    = {Ludmila Viotti and M. Bel{\'e}n Far{\'\i}as and Paula I. Villar and Fernando C. Lombardo},
  title     = {Thermal corrections to quantum friction and decoherence: A closed-time-path approach to atom-surface interaction},
  journal   = {Physical Review D},
  volume    = {99},
  pages     = {105005},
  year      = {2019},
  month     = may,
  publisher = {American Physical Society},
  doi       = {10.1103/PhysRevD.99.105005},
  url       = {https://doi.org/10.1103/PhysRevD.99.105005}
}

@article{Jeske2013,
  author    = {Jan Jeske and Jared H. Cole},
  title     = {Derivation of Markovian master equations for spatially correlated decoherence},
  journal   = {Physical Review A},
  volume    = {87},
  pages     = {052138},
  year      = {2013},
  publisher = {American Physical Society},
  doi       = {10.1103/PhysRevA.87.052138},
  url       = {https://doi.org/10.1103/PhysRevA.87.052138}
}

@article{Wilson2022,
  author  = {Wilson, Kali E. and Samson, E. Carlo and Newman, Zachary L. and Anderson, Brian P.},
  title   = {Generation of high-winding-number superfluid circulation in Bose-Einstein condensates},
  journal = {Physical Review A},
  volume  = {106},
  number  = {3},
  pages   = {033319},
  year    = {2022},
  doi     = {10.1103/PhysRevA.106.033319},
  url     = {https://doi.org/10.1103/PhysRevA.106.033319}
}

@article{OConnell2010,
  author = {O'Connell, A. D. and et al.},
  title = {Quantum Ground State and Single-Phonon Control of a Mechanical Resonator},
  journal = {Nature},
  volume = {464},
  pages = {697--703},
  year = {2010},
  doi = {10.1038/nature08967},
  url = {https://doi.org/10.1038/nature08967}
}

@article{Zou2024SpatiallyCorrelatedNoise,
  author    = {Ji Zou and Stefano Bosco and Daniel Loss},
  title     = {Spatially correlated classical and quantum noise in driven qubits},
  journal   = {npj Quantum Information},
  volume    = {10},
  number    = {1},
  pages     = {46},
  year      = {2024},
  doi       = {10.1038/s41534-024-00842-9},
  url       = {https://doi.org/10.1038/s41534-024-00842-9}
}

@article{Gamba2025DrivenParticleBath,
 title = {Open quantum systems with particle and bath driven by time-dependent fields},
  author = {Gamba, Daniele and Cui, Bingyu and Zaccone, Alessio},
  journal = {Phys. Rev. A},
  volume = {112},
  issue = {1},
  pages = {012207},
  numpages = {9},
  year = {2025},
  month = {Jul},
  publisher = {American Physical Society},
  doi = {10.1103/w3vk-wx62}
}

@article{Leboeuf2001,
  author  = {Leboeuf, P. and Pavloff, N.},
  title   = {Bose-Einstein beams: Coherent propagation through a guide},
  journal = {Physical Review A},
  volume  = {64},
  number  = {3},
  pages   = {033602},
  year    = {2001},
  doi     = {10.1103/PhysRevA.64.033602},
  url     = {https://doi.org/10.1103/PhysRevA.64.033602}
}

@article{Tyagi2024HuygensClock,
  author    = {Bhavay Tyagi and Hao Li and Eric R. Bittner and Andrei Piryatinski and Carlos Silva-Acu{\~n}a},
  title     = {Noise-Induced Quantum Synchronization and Entanglement in a Quantum Analogue of Huygens' Clock},
  journal   = {The Journal of Physical Chemistry Letters},
  volume    = {15},
  number    = {43},
  pages     = {10896--10902},
  year      = {2024},
  doi       = {10.1021/acs.jpclett.4c02313},
  url       = {https://doi.org/10.1021/acs.jpclett.4c02313}
}

@article{Krzywda2026MobileSpinQubits,
  author    = {Jan A. Krzywda and Yuta Matsumoto and Maxim De Smet and Larysa Tryputen and Sander L. de Snoo and Sergey V. Amitonov and Evert van Nieuwenburg and Giordano Scappucci and Lieven M. K. Vandersypen},
  title     = {Coherence Protection for Mobile Spin Qubits in Silicon},
  journal   = {arXiv},
  volume    = {arXiv:2602.09179},
  year      = {2026},
  doi       = {10.48550/arXiv.2602.09179},
  url       = {https://doi.org/10.48550/arXiv.2602.09179},
  note      = {Preprint}
}

@article{Zurek1991vd,
    author = "Zurek, Wojciech H.",
    title = "{Decoherence and the transition from quantum to classical}",
    eprint = "quant-ph/0306072",
    archivePrefix = "arXiv",
    doi = "10.1063/1.881293",
    journal = "Phys. Today",
    volume = {44},
    number = {10},
    pages = "36--44",
    year = "1991"
}

@Inbook{Paz2002,
author="Paz, Juan Pablo
and Zurek, Wojciech Hubert",
editor="Heiss, Dieter",
title="Environment-Induced Decoherence and the Transition from Quantum to Classical",
bookTitle="Fundamentals of Quantum Information: Quantum Computation, Communication, Decoherence and All That",
year="2002",
publisher="Springer Berlin Heidelberg",
address="Berlin, Heidelberg",
pages="77--148",
isbn="978-3-540-45933-0",
doi="10.1007/3-540-45933-2_4"
}

@Inbook{Ingold2002,
author="Ingold, Gert-Ludwig",
editor="Buchleitner, Andreas
and Hornberger, Klaus",
title="Path Integrals and Their Application to Dissipative Quantum Systems",
bookTitle="Coherent Evolution in Noisy Environments",
year="2002",
publisher="Springer Berlin Heidelberg",
address="Berlin, Heidelberg",
pages="1--53",
isbn="978-3-540-45855-5",
doi="10.1007/3-540-45855-7_1"
}

@article{10.1063/1.531046,
    author = {Makri, Nancy},
    title = {Numerical path integral techniques for long time dynamics of quantum dissipative systems},
    journal = {Journal of Mathematical Physics},
    volume = {36},
    number = {5},
    pages = {2430-2457},
    year = {1995},
    month = {05},
    issn = {0022-2488},
    doi = {10.1063/1.531046}
}

@book{banerjee2018open,
  author    = {Banerjee, Subhashish},
  title     = {Open Quantum Systems: Dynamics of Nonclassical Evolution},
  publisher = {Springer},
  address   = {Singapore},
  year      = {2018},
  doi       = {10.1007/978-981-13-3182-4}
}

@article{z3gm32jn,
  title = {Engineering active motion in quantum matter},
  author = {Antonov, Alexander P. and Zheng, Yuanjian and Liebchen, Benno and L\"owen, Hartmut},
  journal = {Phys. Rev. Res.},
  volume = {7},
  issue = {3},
  pages = {033008},
  numpages = {8},
  year = {2025},
  month = {Jul},
  publisher = {American Physical Society},
  doi = {10.1103/z3gm-32jn},
  url = {https://link.aps.org/doi/10.1103/z3gm-32jn}
}

@article{PhysRevA.106.052205,
  title = {Electromagnetic viscosity in complex structured environments: From blackbody to quantum friction},
  author = {Oelschl\"ager, M. and Reiche, D. and Egerland, C. H. and Busch, K. and Intravaia, F.},
  journal = {Phys. Rev. A},
  volume = {106},
  issue = {5},
  pages = {052205},
  numpages = {16},
  year = {2022},
  month = {Nov},
  publisher = {American Physical Society},
  doi = {10.1103/PhysRevA.106.052205},
  url = {https://link.aps.org/doi/10.1103/PhysRevA.106.052205}
}

@article{fskmy179,
  title = {Pseudoratchet through one-dimensional quantum Brownian motion for energy-efficient Brownian computing},
  author = {Nakade, Sho and Peper, Ferdinand and Kanki, Kazuki and Petrosky, Tomio},
  journal = {Phys. Rev. Res.},
  volume = {7},
  issue = {4},
  pages = {043316},
  numpages = {19},
  year = {2025},
  month = {Dec},
  publisher = {American Physical Society},
  doi = {10.1103/fskm-y179},
  url = {https://link.aps.org/doi/10.1103/fskm-y179}
}

@article{RevModPhys.88.041001,
  title = {Colloquium: Protecting quantum information against environmental noise},
  author = {Suter, Dieter and \'Alvarez, Gonzalo A.},
  journal = {Rev. Mod. Phys.},
  volume = {88},
  issue = {4},
  pages = {041001},
  numpages = {23},
  year = {2016},
  month = {Oct},
  publisher = {American Physical Society},
  doi = {10.1103/RevModPhys.88.041001}
}

@article{1609231253819,
  title = {From quantum chemistry to quantum biology: a path toward consciousness},
  author = {Jack A. Tuszynski},
  journal = {J. Integr. Neurosci.},
  volume = {19},
  number = {4},
  pages = {687--700},
  doi = {10.31083/j.jin.2020.04.393},
  year = {2020}

}

@article{Campbell_2026,
doi = {10.1088/2058-9565/ae1e27},
url = {https://doi.org/10.1088/2058-9565/ae1e27},
year = {2026},
month = {jan},
publisher = {IOP Publishing},
volume = {11},
number = {1},
pages = {012501},
title = {Roadmap on quantum thermodynamics},
journal = {Quantum Science and Technology},
author = {Campbell, Steve and et al.}
}

@article{10.1063/5.0197142,
    author = {Balandin, Alexander A. and Paladino, Elisabetta and Hakonen, Pertti J.},
    title = {Electronic noise—From advanced materials to quantum technologies},
    journal = {Applied Physics Letters},
    volume = {124},
    number = {5},
    pages = {050401},
    year = {2024},
    month = {01},
    issn = {0003-6951},
    doi = {10.1063/5.0197142}
}

@article{PhysRevApplied.20.034038,
  title = {Effect of Measurement Backaction on Quantum Clock Precision Studied with a Superconducting Circuit},
  author = {He, Xin and Pakkiam, Prasanna and Gangat, Adil A. and Kewming, Michael J. and Milburn, Gerard J. and Fedorov, Arkady},
  journal = {Phys. Rev. Appl.},
  volume = {20},
  issue = {3},
  pages = {034038},
  numpages = {17},
  year = {2023},
  month = {Sep},
  publisher = {American Physical Society},
  doi = {10.1103/PhysRevApplied.20.034038},
  url = {https://link.aps.org/doi/10.1103/PhysRevApplied.20.034038}
}

@article{Liu_2022,
  author = {Liu, Yulong and Zhou, Jingwei and Mercier de L{\'e}pinay, Laure and Sillanp{\"a}{\"a}, Mika A.},
  title = {Quantum backaction evading measurements of a silicon nitride membrane resonator},
  journal = {New Journal of Physics},
  volume = {24},
  number = {8},
  pages = {083043},
  year = {2022},
  month = {aug},
  publisher = {IOP Publishing},
  doi = {10.1088/1367-2630/ac88ef},
  url = {https://doi.org/10.1088/1367-2630/ac88ef}
}

@article{harrington2022,
  title={Engineered dissipation for quantum information science},
  author={Harrington, Patrick M and Mueller, Erich J and Murch, Kater W},
  journal={Nature Reviews Physics},
  volume={4},
  number={10},
  pages={660--671},
  year={2022},
  publisher={Nature Publishing Group UK London},
  doi = {10.1038/s42254-022-00494-8}
}

@ARTICLE{Shrikant2023,
AUTHOR={Shrikant, U.  and Mandayam, Prabha },       
TITLE={Quantum non-Markovianity: Overview and recent developments},      
JOURNAL={Frontiers in Quantum Science and Technology},      
volume = {2},
pages  = {1134583},
YEAR={2023},
DOI={10.3389/frqst.2023.1134583},
ISSN={2813-2181},
}

@article{Rotter_2015,
doi = {10.1088/0034-4885/78/11/114001},
year = {2015},
month = {oct},
publisher = {IOP Publishing},
volume = {78},
number = {11},
pages = {114001},
author = {Rotter, I and Bird, J P},
title = {A review of progress in the physics of open quantum systems: theory and experiment},
journal = {Reports on Progress in Physics}
}

@article{PhysRevA.106.032220,
  title = {Markovian and non-Markovian dynamics of quantum coherence in the extended $XX$ chain},
  author = {Yin, Shaoying and Liu, Shutian and Song, Jie and Luan, Hongliang},
  journal = {Phys. Rev. A},
  volume = {106},
  issue = {3},
  pages = {032220},
  numpages = {8},
  year = {2022},
  month = {Sep},
  publisher = {American Physical Society},
  doi = {10.1103/PhysRevA.106.032220},
  url = {https://link.aps.org/doi/10.1103/PhysRevA.106.032220}
}

@article{PhysRevResearch.7.L012068,
  title = {Non-Markovian skin effect},
  author = {Kuo, Po-Chen and Yang, Shen-Liang and Lambert, Neill and Lin, Jhen-Dong and Huang, Yi-Te and Nori, Franco and Chen, Yueh-Nan},
  journal = {Phys. Rev. Res.},
  volume = {7},
  issue = {1},
  pages = {L012068},
  numpages = {7},
  year = {2025},
  month = {Mar},
  publisher = {American Physical Society},
  doi = {10.1103/PhysRevResearch.7.L012068},
  url = {https://link.aps.org/doi/10.1103/PhysRevResearch.7.L012068}
}

@article{PhysRevA.101.013822,
  title = {Coherence-based measurement of non-Markovian dynamics in an open quantum system},
  author = {Yugra, Yonny and De Zela, Francisco and Cuevas, \'Alvaro},
  journal = {Phys. Rev. A},
  volume = {101},
  issue = {1},
  pages = {013822},
  numpages = {8},
  year = {2020},
  month = {Jan},
  publisher = {American Physical Society},
  doi = {10.1103/PhysRevA.101.013822},
  url = {https://link.aps.org/doi/10.1103/PhysRevA.101.013822}
}

@article{Wu4ltm,
  title = {Steady-state coherences under partial collective non-Markovian decoherence},
  author = {Wu, S. L. and Ma, W. and Wang, Zhao-Ming and Brumer, P. and Wu, Lian-Ao},
  journal = {Phys. Rev. Res.},
  volume = {7},
  issue = {4},
  pages = {043221},
  numpages = {15},
  year = {2025},
  month = {Nov},
  publisher = {American Physical Society},
  doi = {10.1103/nktx-4ltm},
  url = {https://link.aps.org/doi/10.1103/nktx-4ltm}
}

@article{PhysRevA.110.052220,
  title = {Non-Markovian-environment-induced anomaly in steady-state quantum coherence},
  author = {Ablimit, Arapat and Wang, Zhao-Ming and Ren, Feng-Hua and Brumer, Paul and Wu, Lian-Ao},
  journal = {Phys. Rev. A},
  volume = {110},
  issue = {5},
  pages = {052220},
  numpages = {7},
  year = {2024},
  month = {Nov},
  publisher = {American Physical Society},
  doi = {10.1103/PhysRevA.110.052220},
  url = {https://link.aps.org/doi/10.1103/PhysRevA.110.052220}
}

@article{PhysRevA.105.012209,
  title = {Master equation for non-Markovian quantum Brownian motion: The emergence of lateral coherences},
  author = {Lally, Sapphire and Werren, Nicholas and Al-Khalili, Jim and Rocco, Andrea},
  journal = {Phys. Rev. A},
  volume = {105},
  issue = {1},
  pages = {012209},
  numpages = {16},
  year = {2022},
  month = {Jan},
  publisher = {American Physical Society},
  doi = {10.1103/PhysRevA.105.012209},
  url = {https://link.aps.org/doi/10.1103/PhysRevA.105.012209}
}

@article{zvkls7hy,
  title = {Electronic memory of vibrational dynamics: Non-Markovian evolution of electronic coherence from complex dynamical weak values},
  author = {Vatasescu, Mihaela},
  journal = {Phys. Rev. A},
  volume = {113},
  issue = {1},
  pages = {012404},
  numpages = {15},
  year = {2026},
  month = {Jan},
  publisher = {American Physical Society},
  doi = {10.1103/zvkl-s7hy},
  url = {https://link.aps.org/doi/10.1103/zvkl-s7hy}
}

@article{PhysRevA.94.042110,
  title = {Decoherence induced by non-Markovian noise in a nonequilibrium environment},
  author = {Cai, Xiangji and Zheng, Yujun},
  journal = {Phys. Rev. A},
  volume = {94},
  issue = {4},
  pages = {042110},
  numpages = {5},
  year = {2016},
  month = {Oct},
  publisher = {American Physical Society},
  doi = {10.1103/PhysRevA.94.042110},
  url = {https://link.aps.org/doi/10.1103/PhysRevA.94.042110}
}

@article{10.1063/5.0083067,
    author = {Reiche, D. and Intravaia, F. and Busch, K.},
    title = {Wading through the void: Exploring quantum friction and nonequilibrium fluctuations},
    journal = {APL Photonics},
    volume = {7},
    number = {3},
    pages = {030902},
    year = {2022},
    month = {03},
    issn = {2378-0967},
    doi = {10.1063/5.0083067},
    url = {https://doi.org/10.1063/5.0083067}
}

@article{Sarkar97h,
  title = {Non-Markovian route to coherence in heterogeneous diffusive systems},
  author = {Sarkar, Aranyak},
  journal = {Phys. Rev. E},
  volume = {112},
  issue = {5},
  pages = {054117},
  numpages = {7},
  year = {2025},
  month = {Nov},
  publisher = {American Physical Society},
  doi = {10.1103/6r83-n97h},
  url = {https://link.aps.org/doi/10.1103/6r83-n97h}
}

@article{doi:10.1142/S0129055X03001631,
author = {BLANCHARD, PH. and OLKIEWICZ, R.},
title = {DECOHERENCE INDUCED TRANSITION FROM QUANTUM TO CLASSICAL DYNAMICS},
journal = {Reviews in Mathematical Physics},
volume = {15},
number = {03},
pages = {217-243},
year = {2003},
doi = {10.1142/S0129055X03001631}
}

@article{RevModPhys.75.715,
  title = {Decoherence, einselection, and the quantum origins of the classical},
  author = {Zurek, Wojciech Hubert},
  journal = {Rev. Mod. Phys.},
  volume = {75},
  issue = {3},
  pages = {715--775},
  numpages = {0},
  year = {2003},
  month = {May},
  publisher = {American Physical Society},
  doi = {10.1103/RevModPhys.75.715},
  url = {https://link.aps.org/doi/10.1103/RevModPhys.75.715}
}

@article{PhysRevLett.104.200401,
  title = {Sudden Transition between Classical and Quantum Decoherence},
  author = {Mazzola, L. and Piilo, J. and Maniscalco, S.},
  journal = {Phys. Rev. Lett.},
  volume = {104},
  issue = {20},
  pages = {200401},
  numpages = {4},
  year = {2010},
  month = {May},
  publisher = {American Physical Society},
  doi = {10.1103/PhysRevLett.104.200401},
  url = {https://link.aps.org/doi/10.1103/PhysRevLett.104.200401}
}

@article{Caldeira1983,
  author = {Caldeira, A. O. and Leggett, A. J.},
  title = {Path Integral Approach to Quantum Brownian Motion},
  journal = {Physica A},
  volume = {121},
  number = {3},
  pages = {587--616},
  year = {1983},
  doi = {10.1016/0378-4371(83)90013-4}
}

@article{Leggett1987,
  author = {Leggett, A. J. and Chakravarty, S. and Dorsey, A. T. and Fisher, M. P. A. and Garg, A. and Zwerger, W.},
  title = {Dynamics of the Dissipative Two-State System},
  journal = {Reviews of Modern Physics},
  volume = {59},
  number = {1},
  pages = {1--85},
  year = {1987},
  doi = {10.1103/RevModPhys.59.1}
}

@article{Braun2002,
  author = {Braun, Daniel},
  title = {Creation of Entanglement by Interaction with a Common Heat Bath},
  journal = {Physical Review Letters},
  volume = {89},
  number = {27},
  pages = {277901},
  year = {2002},
  doi = {10.1103/PhysRevLett.89.277901}
}

@article{Brattegard2024,
  title = {Thermometry by correlated dephasing of impurities in a one-dimensional Fermi gas},
  author = {Brattegard, Sindre and Mitchison, Mark T.},
  journal = {Phys. Rev. A},
  volume = {109},
  issue = {2},
  pages = {023309},
  numpages = {15},
  year = {2024},
  month = {Feb},
  publisher = {American Physical Society},
  doi = {10.1103/PhysRevA.109.023309}
}

@article{PhysRevResearch.6.043222,
  title = {Avoiding decoherence with giant atoms in a two-dimensional structured environment},
  author = {Raaholt Ingelsten, Emil and Kockum, Anton Frisk and Soro, Ariadna},
  journal = {Phys. Rev. Res.},
  volume = {6},
  issue = {4},
  pages = {043222},
  numpages = {20},
  year = {2024},
  month = {Dec},
  publisher = {American Physical Society},
  doi = {10.1103/PhysRevResearch.6.043222},
  url = {https://link.aps.org/doi/10.1103/PhysRevResearch.6.043222}
}

@article{Feynman1963,
  author = {Feynman, R. P. and Vernon, F. L.},
  title = {The Theory of a General Quantum System Interacting with a Linear Dissipative System},
  journal = {Annals of Physics},
  volume = {24},
  number = {1},
  pages = {118--173},
  year = {1963},
  doi = {10.1016/0003-4916(63)90068-X}
}

@book{Calzetta2008,
  author = {Calzetta, Esteban A. and Hu, Bei-Lok B.},
  title = {Nonequilibrium Quantum Field Theory},
  publisher = {Cambridge University Press},
  address = {Cambridge},
  year = {2008},
  doi = {10.1017/CBO9780511535123}
}

@article{Sieberer2016,
  author = {Sieberer, L. M. and Buchhold, M. and Diehl, S.},
  title = {Keldysh Field Theory for Driven Open Quantum Systems},
  journal = {Reports on Progress in Physics},
  volume = {79},
  number = {9},
  pages = {096001},
  year = {2016},
  doi = {10.1088/0034-4885/79/9/096001}
}

@article{sung2019non,
  title={Non-Gaussian noise spectroscopy with a superconducting qubit sensor},
  author={Sung, Youngkyu and Beaudoin, F{\'e}lix and Norris, Leigh M and Yan, Fei and Kim, David K and Qiu, Jack Y and von L{\"u}pke, Uwe and Yoder, Jonilyn L and Orlando, Terry P and Gustavsson, Simon and others},
  journal={Nature communications},
  volume={10},
  number={1},
  pages={3715},
  year={2019},
  publisher={Nature Publishing Group UK London},
  doi = {10.1038/s41467-019-11699-4}
}

@article{PhysRevResearch.7.023073,
  title = {Temporally correlated quantum noise in driven quantum systems with applications to quantum gate operations},
  author = {Gul\'acsi, Bal\'azs and Burkard, Guido},
  journal = {Phys. Rev. Res.},
  volume = {7},
  issue = {2},
  pages = {023073},
  numpages = {17},
  year = {2025},
  month = {Apr},
  publisher = {American Physical Society},
  doi = {10.1103/PhysRevResearch.7.023073},
  url = {https://link.aps.org/doi/10.1103/PhysRevResearch.7.023073}
}

@article{Berges2004,
  author = {Berges, J.},
  title = {Introduction to Nonequilibrium QFT},
  journal = {AIP Conf. Proc.},
  volume = {739},
  pages = {3},
  year = {2004},
  doi = {10.1063/1.1843591}
}

@book{Altland2010,
  author = {Altland, A. and Simons, B.},
  title = {Condensed Matter Field Theory},
  publisher = {Cambridge University Press},
  year = {2010},
  doi = {10.1017/CBO9780511789984}
}

@book{Kamenev2011,
  author = {Kamenev, Alex},
  title = {Field Theory of Non-Equilibrium Systems},
  publisher = {Cambridge University Press},
  address = {Cambridge},
  year = {2011},
  doi = {10.1017/CBO9781139003667}
}

@article{Reina2002,
  author = {Reina, John H. and Quiroga, Luis and Johnson, Neil F.},
  title = {Decoherence of Quantum Registers},
  journal = {Physical Review A},
  volume = {65},
  number = {3},
  pages = {032326},
  year = {2002},
  doi = {10.1103/PhysRevA.65.032326}
}

@article{Strathearn2018,
  author  = {Strathearn, A. and Kirton, P. and Kilda, D.
             and Keeling, J. and Lovett, B. W.},
  title   = {Efficient Non-Markovian Quantum Dynamics Using
             Time-Evolving Matrix Product Operators},
  journal = {Nature Communications},
  volume  = {9},
  pages   = {3322},
  year    = {2018},
  doi     = {10.1038/s41467-018-05617-3}
}

@article{Pendry1997,
  doi = {10.1088/0953-8984/9/47/001},
  year = {1997},
  month = {nov},
  volume = {9},
  number = {47},
  pages = {10301},
  author = {J B Pendry},
  title = {Shearing the vacuum - quantum friction},
  journal = {Journal of Physics: Condensed Matter}
}

@article{Volokitin2007,
  author = {Volokitin, A. I. and Persson, B. N. J.},
  title = {Near-Field Radiative Heat Transfer and Noncontact Friction},
  journal = {Reviews of Modern Physics},
  volume = {79},
  number = {4},
  pages = {1291--1329},
  year = {2007},
  doi = {10.1103/RevModPhys.79.1291}
}

@article{Kirton2020,
  author = {Kirton, Peter and Roses, Matthew M. and Keeling, Jonathan and Dalla Torre, Emanuele G.},
  title = {Introduction to the Dicke Model. From Equilibrium to Nonequilibrium, and Vice Versa},
  journal = {Advanced Quantum Technologies},
  volume = {2},
  number = {1-2},
  pages = {1800043},
  year = {2019},
  doi = {10.1002/qute.201800043}
}

@article{McCutcheon2011,
  author = {McCutcheon, Dara P. S. and Dattani, Nikesh S. and Gauger, Erik M. and Lovett, Brendon W. and Nazir, Ahsan},
  title = {A General Approach to Quantum Dynamics Using a Variational Master Equation. Application to Phonon-Damped Rabi Rotations in Quantum Dots},
  journal = {Physical Review B},
  volume = {84},
  number = {8},
  pages = {081305},
  year = {2011},
  doi = {10.1103/PhysRevB.84.081305}
}

@article{Chin2013,
  author  = {Chin, A. W. and Prior, J. and Rosenbach, R.
             and Caycedo-Soler, F. and Huelga, S. F.
             and Plenio, Martin B.},
  title   = {The Role of Non-Equilibrium Vibrational Structures in
             Electronic Coherence and Recoherence in Pigment-Protein
             Complexes},
  journal = {Nature Physics},
  volume  = {9},
  pages   = {113--118},
  year    = {2013},
  doi     = {10.1038/nphys2515}
}

@article{Tamascelli2019,
  author = {Tamascelli, Dario and Smirne, Andrea and Lim, James and Huelga, Susana F. and Plenio, Martin B.},
  title = {Efficient Simulation of Finite-Temperature Open Quantum Systems},
  journal = {Physical Review Letters},
  volume = {123},
  number = {9},
  pages = {090402},
  year = {2019},
  doi = {10.1103/PhysRevLett.123.090402}
}

@article{Schuetz2015,
 title = {Universal Quantum Transducers Based on Surface Acoustic Waves},
  author = {Schuetz, M. J. A. and Kessler, E. M. and Giedke, G. and Vandersypen, L. M. K. and Lukin, M. D. and Cirac, J. I.},
  journal = {Phys. Rev. X},
  volume = {5},
  issue = {3},
  pages = {031031},
  numpages = {30},
  year = {2015},
  month = {Sep},
  publisher = {American Physical Society},
  doi = {10.1103/PhysRevX.5.031031}
}

@article{Hu1992,
  author  = {Hu, B. L. and Paz, J. P. and Zhang, Y.},
  title   = {Quantum Brownian Motion in a General Environment:
             Exact Master Equation with Nonlocal Dissipation and Colored Noise},
  journal = {Physical Review D},
  volume  = {45},
  number  = {8},
  pages   = {2843--2861},
  year    = {1992},
  doi     = {10.1103/PhysRevD.45.2843},
  url     = {https://doi.org/10.1103/PhysRevD.45.2843}
}

@book{Breuer2002,
  author    = {Breuer, Heinz-Peter and Petruccione, Francesco},
  title     = {The Theory of Open Quantum Systems},
  publisher = {Oxford University Press},
  address   = {Oxford},
  year      = {2007},
  doi       = {10.1093/acprof:oso/9780199213900.001.0001}
}

@book{Weiss2012,
  author = {Weiss, Ulrich},
  title = {Quantum Dissipative Systems},
  edition = {4},
  publisher = {World Scientific},
  address = {Singapore},
  year = {2012},
  doi = {10.1142/8334}
}

@book{Schlosshauer2007,
title="The Basic Formalism and Interpretation of Decoherence",
bookTitle="Decoherence and the Quantum-To-Classical Transition",
author="M Schlosshauer",
year="2007",
publisher="Springer Berlin Heidelberg",
address="Berlin, Heidelberg",
pages="13--114",
doi="10.1007/978-3-540-35775-9_2"
}

@book{Carmichael1999,
  author = {Carmichael, Howard J.},
  title = {Statistical Methods in Quantum Optics 1. Master Equations and Fokker-Planck Equations},
  publisher = {Springer},
  address = {Berlin Heidelberg},
  year = {1999},
  doi = {10.1007/978-3-662-03875-8}
}

@article{Clerk2010,
  author = {Clerk, A. A. and Devoret, M. H. and Girvin, S. M. and Marquardt, F. and Schoelkopf, R. J.},
  title = {Introduction to Quantum Noise, Measurement, and Amplification},
  journal = {Reviews of Modern Physics},
  volume = {82},
  number = {2},
  pages = {1155--1208},
  year = {2010},
  doi = {10.1103/RevModPhys.82.1155}
}

@book{Rivas2012,
  author = {Rivas, {\'A}ngel and Huelga, Susana F.},
  title = {Open Quantum Systems. An Introduction},
  publisher = {Springer},
  address = {Heidelberg},
  year = {2012},
  doi = {10.1007/978-3-642-23354-8}
}

@article{DeVega2017,
  author = {de Vega, In{\'e}s and Alonso, Daniel},
  title = {Dynamics of Non-Markovian Open Quantum Systems},
  journal = {Reviews of Modern Physics},
  volume = {89},
  number = {1},
  pages = {015001},
  year = {2017},
  doi = {10.1103/RevModPhys.89.015001}
}

@article{Chenu2020,
  author = {Chenu, Aur{\'e}lia and Beau, Mathieu and Cao, Jianshu and del Campo, Adolfo},
  title = {Quantum Simulation of Generic Many-Body Open System Dynamics Using Classical Noise},
  journal = {Physical Review Letters},
  volume = {118},
  number = {14},
  pages = {140403},
  year = {2017},
  doi = {10.1103/PhysRevLett.118.140403}
}

@article{Maghrebi2015,
  title = {Nonequilibrium quantum fluctuations of a dispersive medium: Spontaneous emission, photon statistics, entropy generation, and stochastic motion},
  author = {Maghrebi, Mohammad F. and Jaffe, Robert L. and Kardar, Mehran},
  journal = {Phys. Rev. A},
  volume = {90},
  issue = {1},
  pages = {012515},
  numpages = {22},
  year = {2014},
  month = {Jul},
  publisher = {American Physical Society},
  doi = {10.1103/PhysRevA.90.012515}
}

@article{Strunz1999,
  author = {Strunz, Walter T. and Di{\'o}si, Lajos and Gisin, Nicolas},
  title = {Open System Dynamics with Non-Markovian Quantum Trajectories},
  journal = {Physical Review Letters},
  volume = {82},
  number = {9},
  pages = {1801--1805},
  year = {1999},
  doi = {10.1103/PhysRevLett.82.1801}
}

@article{Lahteenmaki2013,
  author = {L{\"a}hteenm{\"a}ki, P. and Paraoanu, G. S. and Hassel, J. and Hakonen, P. J.},
  title = {Dynamical Casimir Effect in a Josephson Metamaterial},
  journal = {Proceedings of the National Academy of Sciences},
  volume = {110},
  number = {11},
  pages = {4234--4238},
  year = {2013},
  doi = {10.1073/pnas.1212705110}
}

@article{Gustafsson2014,
  author = {Gustafsson, M. V. and Aref, T. and Kockum, A. F. and Ekstr{\"o}m, M. K. and Johansson, G. and Delsing, P.},
  title = {Propagating Phonons Coupled to an Artificial Atom},
  journal = {Science},
  volume = {346},
  number = {6206},
  pages = {207--211},
  year = {2014},
  doi = {10.1126/science.1257219}
}

@article{Satzinger2018,
  author  = {Satzinger, K. J. and Zhong, Y. P. and Chang, H.-S.
             and Peairs, G. A. and Bienfait, A. and Chou, Ming-Han
             and Cleland, A. Y. and Conner, C. R. and Dumur, {\'E}.
             and Grebel, J. and Gutierrez, I. and November, B. H.
             and Povey, R. G. and Whiteley, S. J. and Awschalom, D. D.
             and Schuster, D. I. and Cleland, A. N.},
  title   = {Quantum Control of Surface Acoustic-Wave Phonons},
  journal = {Nature},
  volume  = {563},
  pages   = {661--665},
  year    = {2018},
  doi     = {10.1038/s41586-018-0719-5}
}

@article{Kurpiers2018,
  author  = {Kurpiers, P. and Magnard, P. and Walter, T. and Royer, B.
             and Pechal, M. and Heinsoo, J. and Salath{\'e}, Y.
             and Akin, A. and Storz, S. and Besse, J.-C.
             and Gasparinetti, S. and Blais, A. and Wallraff, A.},
  title   = {Deterministic Quantum State Transfer and Remote Entanglement
             Using Microwave Photons},
  journal = {Nature},
  volume  = {558},
  pages   = {264--267},
  year    = {2018},
  doi     = {10.1038/s41586-018-0195-y}
}

@article{vonLupke2020,
  title = {Two-Qubit Spectroscopy of Spatiotemporally Correlated Quantum Noise in Superconducting Qubits},
  author = {von L\"upke, Uwe and Beaudoin, F\'elix and Norris, Leigh M. and Sung, Youngkyu and Winik, Roni and Qiu, Jack Y. and Kjaergaard, Morten and Kim, David and Yoder, Jonilyn and Gustavsson, Simon and Viola, Lorenza and Oliver, William D.},
  journal = {PRX Quantum},
  volume = {1},
  issue = {1},
  pages = {010305},
  numpages = {23},
  year = {2020},
  month = {Sep},
  publisher = {American Physical Society},
  doi = {10.1103/PRXQuantum.1.010305}
}

@article{Engels2007,
  title = {Stationary and Nonstationary Fluid Flow of a Bose-Einstein Condensate Through a Penetrable Barrier},
  author = {Engels, P. and Atherton, C.},
  journal = {Phys. Rev. Lett.},
  volume = {99},
  pages = {160405},
  year = {2007},
  doi = {10.1103/PhysRevLett.99.160405},
  url = {https://doi.org/10.1103/PhysRevLett.99.160405}
}

@article{Desbuquois2012,
  title = {Superfluid behaviour of a two-dimensional Bose gas},
  author = {Desbuquois, R. and et al.},
  journal = {Nat. Phys.},
  volume = {8},
  pages = {645--648},
  year = {2012},
  doi = {10.1038/nphys2378},
  url = {https://doi.org/10.1038/nphys2378}
}

@article{Basov2016,
  author  = {Basov, D. N. and Fogler, M. M. and Garc{\'i}a de Abajo, F. J.},
  title   = {Polaritons in van der Waals materials},
  journal = {Science},
  volume  = {354},
  number  = {6309},
  pages   = {aag1992},
  year    = {2016},
  doi     = {10.1126/science.aag1992},
  url     = {https://doi.org/10.1126/science.aag1992}
}

@article{Morgado2018,
  author  = {Morgado, Tiago A. and Silveirinha, M{\'a}rio G.},
  title   = {Drift-Induced Unidirectional Graphene Plasmons},
  journal = {ACS Photonics},
  volume  = {5},
  number  = {11},
  pages   = {4253--4258},
  year    = {2018},
  doi     = {10.1021/acsphotonics.8b00987},
  url     = {https://doi.org/10.1021/acsphotonics.8b00987}
}

@article{In2022,
  author  = {In, Chihun and Kim, Un Jeong and Choi, Hyunyong},
  title   = {Two-dimensional Dirac plasmon-polaritons in graphene, 3D topological insulator and hybrid systems},
  journal = {Light: Science \& Applications},
  volume  = {11},
  pages   = {313},
  year    = {2022},
  doi     = {10.1038/s41377-022-01012-2},
  url     = {https://doi.org/10.1038/s41377-022-01012-2}
}

@article{Goldflam2015,
  author  = {Goldflam, Michael D. and others},
  title   = {Tuning and Persistent Switching of Graphene Plasmons on a Ferroelectric Substrate},
  journal = {Nano Letters},
  volume  = {15},
  number  = {8},
  pages   = {4859--4864},
  year    = {2015},
  doi     = {10.1021/acs.nanolett.5b00125},
  url     = {https://doi.org/10.1021/acs.nanolett.5b00125}
}

@article{Carbotte2012,
  author  = {Carbotte, J. P. and LeBlanc, J. P. F. and Nicol, E. J.},
  title   = {Emergence of plasmaronic structure in the near-field optical response of graphene},
  journal = {Physical Review B},
  volume  = {85},
  number  = {20},
  pages   = {201411},
  year    = {2012},
  doi     = {10.1103/PhysRevB.85.201411},
  url     = {https://doi.org/10.1103/PhysRevB.85.201411}
}

@article{Wagner2014,
  author  = {Wagner, Martin and others},
  title   = {Ultrafast and nanoscale plasmonic phenomena in exfoliated graphene revealed by infrared pump-probe nanoscopy},
  journal = {Nano Letters},
  volume  = {14},
  number  = {2},
  pages   = {894--900},
  year    = {2014},
  doi     = {10.1021/nl4042577},
  url     = {https://doi.org/10.1021/nl4042577}
}

@article{Anglhuber2025,
  author  = {Anglhuber, Simon and others},
  title   = {Spacetime Imaging of Group and Phase Velocities of Terahertz Surface Plasmon Polaritons in Graphene},
  journal = {Nano Letters},
  volume  = {25},
  pages   = {2125--2132},
  year    = {2025},
  doi     = {10.1021/acs.nanolett.4c04615},
  url     = {https://doi.org/10.1021/acs.nanolett.4c04615}
}

@article{Fei2012,
  author  = {Fei, Z. and Rodin, A. S. and Andreev, G. O. and Bao, W. and McLeod, A. S. and Wagner, M. and Zhang, L. M. and Zhao, Z. and Dominguez, G. and Thiemens, M. and Fogler, M. M. and Castro Neto, A. H. and Lau, C. N. and Keilmann, F. and Basov, D. N.},
  title   = {Gate-tuning of graphene plasmons revealed by infrared nano-imaging},
  journal = {Nature},
  volume  = {487},
  number  = {7405},
  pages   = {82--85},
  year    = {2012},
  doi     = {10.1038/nature11253},
  url     = {https://doi.org/10.1038/nature11253}
}

@article{Chen2012,
  author  = {Chen, J. and Badioli, M. and Alonso-Gonz{\'a}lez, P. and Thongrattanasiri, S. and Huth, F. and Osmond, J. and Spasenovi{\'c}, M. and Centeno, A. and Pesquera, A. and Godignon, P. and Elorza, A. Z. and Camara, N. and Garc{\'i}a de Abajo, F. J. and Hillenbrand, R. and Koppens, F. H. L.},
  title   = {Optical nano-imaging of gate-tunable graphene plasmons},
  journal = {Nature},
  volume  = {487},
  number  = {7405},
  pages   = {77--81},
  year    = {2012},
  doi     = {10.1038/nature11254},
  url     = {https://doi.org/10.1038/nature11254}
}

@article{Alonso2017,
  title = {Acoustic terahertz graphene plasmons revealed by photocurrent nanoscopy},
  author = {Alonso-Gonzalez, P. and et al.},
  journal = {Nature Nanotechnology},
  volume = {12},
  pages = {31},
  year = {2017},
  doi = {10.1038/nnano.2016.185},
  url = {https://doi.org/10.1038/nnano.2016.185}
}

@article{Raman1999,
  title = {Evidence for a Critical Velocity in a Bose-Einstein Condensed Gas},
  author = {Raman, C. and K\"ohl, M. and Onofrio, R. and Durfee, D. S. and Kuklewicz, C. E. and Hadzibabic, Z. and Ketterle, W.},
  journal = {Phys. Rev. Lett.},
  volume = {83},
  issue = {13},
  pages = {2502--2505},
  numpages = {0},
  year = {1999},
  month = {Sep},
  publisher = {American Physical Society},
  doi = {10.1103/PhysRevLett.83.2502}
}

@article{Onofrio2000,
  title = {Observation of Superfluid Flow in a Bose-Einstein Condensed Gas},
  author = {Onofrio, R. and et al.},
  journal = {Phys. Rev. Lett.},
  volume = {85},
  pages = {2228},
  year = {2000},
  doi = {10.1103/PhysRevLett.85.2228},
  url = {https://doi.org/10.1103/PhysRevLett.85.2228}
}

@article{Wilson2011,
  author  = {Wilson, C. M. and Johansson, G. and Pourkabirian, A. and Simoen, M. and Johansson, J. R. and Duty, T. and Nori, F. and Delsing, P.},
  title   = {Observation of the Dynamical Casimir Effect in a Superconducting Circuit},
  journal = {Nature},
  volume  = {479},
  number  = {7373},
  pages   = {376--379},
  year    = {2011},
  doi     = {10.1038/nature10561},
  url     = {https://doi.org/10.1038/nature10561}
}

@article{Chu2017,
  author  = {Chu, Yiwen and Kharel, Prashanta and Renninger, William H. and Burkhart, Luke D. and Frunzio, Luigi and Rakich, Peter T. and Schoelkopf, Robert J.},
  title   = {Quantum Acoustics with Superconducting Qubits},
  journal = {Science},
  volume  = {358},
  number  = {6360},
  pages   = {199--202},
  year    = {2017},
  doi     = {10.1126/science.aao1511},
  url     = {https://doi.org/10.1126/science.aao1511}
}

@article{Manenti2017,
  author  = {Manenti, Riccardo and Kockum, Anton F. and Patterson, Andrew and Behrle, Tanja and Rahamim, Joseph and Tancredi, Giovanna and Nori, Franco and Leek, Peter J.},
  title   = {Circuit Quantum Acoustodynamics with Surface Acoustic Waves},
  journal = {Nature Communications},
  volume  = {8},
  pages   = {975},
  year    = {2017},
  doi     = {10.1038/s41467-017-01063-9},
  url     = {https://doi.org/10.1038/s41467-017-01063-9}
}

@article{Elze1987,
  author  = {Elze, Hans-Thomas and Miller, David E. and Redlich, Krzysztof},
  title   = {Gauge theories at finite temperature and chemical potential},
  journal = {Physical Review D},
  volume  = {35},
  number  = {2},
  pages   = {748},
  year    = {1987},
  doi     = {10.1103/PhysRevD.35.748},
  url     = {https://doi.org/10.1103/PhysRevD.35.748}
}

@article{VanDuppen2016,
  author  = {Van Duppen, Ben and Tomadin, Andrea and Grigorenko, Alexander N. and Polini, Marco},
  title   = {Current-induced birefringent absorption and non-reciprocal plasmons in graphene},
  journal = {2D Materials},
  volume  = {3},
  number  = {1},
  pages   = {015011},
  year    = {2016},
  doi     = {10.1088/2053-1583/3/1/015011},
  url     = {https://doi.org/10.1088/2053-1583/3/1/015011}
}

@article{Anderson1963,
  author  = {Anderson, P. W.},
  title   = {Plasmons, Gauge Invariance, and Mass},
  journal = {Physical Review},
  volume  = {130},
  number  = {1},
  pages   = {439--442},
  year    = {1963},
  doi     = {10.1103/PhysRev.130.439},
  url     = {https://doi.org/10.1103/PhysRev.130.439}
}

@article{Gold2002,
  author  = {Gold, A.},
  title   = {Screening behavior of a charged Bose-Einstein condensate including many-body effects},
  journal = {Physical Review B},
  volume  = {65},
  number  = {13},
  pages   = {134521},
  year    = {2002},
  doi     = {10.1103/PhysRevB.65.134521},
  url     = {https://doi.org/10.1103/PhysRevB.65.134521}
}

@article{Davoudi2005,
  author  = {Davoudi, B. and Tosi, M. P.},
  title   = {Single-particle and collective excitations in a charged Bose gas at finite temperature},
  journal = {Physical Review B},
  volume  = {72},
  number  = {13},
  pages   = {134520},
  year    = {2005},
  doi     = {10.1103/PhysRevB.72.134520},
  url     = {https://doi.org/10.1103/PhysRevB.72.134520}
}

@article{Kitzman2023,
  author = {Kitzman, J. M. and Lane, J. R. and Undershute, C. and Harrington, P. M. and Beysengulov, N. R. and Mikolas, C. A. and Murch, K. W. and Pollanen, J.},
  title = {Phononic bath engineering of a superconducting qubit},
  journal = {Nature Communications},
  volume = {14},
  pages = {3910},
  year = {2023},
  doi = {10.1038/s41467-023-39682-0},
  url = {https://doi.org/10.1038/s41467-023-39682-0}
}

@article{Intravaia2014,
  title = {Quantum friction and fluctuation theorems},
  author = {Intravaia, F. and Behunin, R. O. and Dalvit, D. A. R.},
  journal = {Phys. Rev. A},
  volume = {89},
  issue = {5},
  pages = {050101(R)},
  numpages = {5},
  year = {2014},
  month = {May},
  publisher = {American Physical Society},
  doi = {10.1103/PhysRevA.89.050101}
}

@article{Millen2020,
doi = {10.1088/1367-2630/18/1/011002},
year = {2016},
month = {jan},
publisher = {IOP Publishing},
volume = {18},
number = {1},
pages = {011002},
author = {Millen, James and Xuereb, André},
title = {Perspective on quantum thermodynamics},
journal = {New Journal of Physics}
}

@article{Fosco2007,
  author  = {Fosco, C. D. and Lombardo, F. C. and Mazzitelli, F. D.},
  title   = {Quantum dissipative effects in moving mirrors: A functional approach},
  journal = {Physical Review D},
  volume  = {76},
  pages   = {085007},
  year    = {2007},
  doi     = {10.1103/PhysRevD.76.085007},
  url     = {https://doi.org/10.1103/PhysRevD.76.085007}
}

@article{Fosco2008,
  author  = {Fosco, C. D. and Lombardo, F. C. and Mazzitelli, F. D.},
  title   = {Casimir effect with dynamical matter on thin mirrors},
  journal = {Physics Letters B},
  volume  = {669},
  pages   = {371--375},
  year    = {2008},
  doi     = {10.1016/j.physletb.2008.10.004},
  url     = {https://doi.org/10.1016/j.physletb.2008.10.004}
}

@article{Fosco2011,
  author  = {Fosco, C. D. and Lombardo, F. C. and Mazzitelli, F. D.},
  title   = {Quantum dissipative effects in moving imperfect mirrors: Sidewise and normal motions},
  journal = {Physical Review D},
  volume  = {84},
  pages   = {025011},
  year    = {2011},
  doi     = {10.1103/PhysRevD.84.025011},
  url     = {https://doi.org/10.1103/PhysRevD.84.025011}
}

@article{Maghrebi2013,
  author  = {Maghrebi, Mohammad F. and Golestanian, Ramin and Kardar, Mehran},
  title   = {Quantum {Cherenkov} radiation and noncontact friction},
  journal = {Physical Review A},
  volume  = {88},
  pages   = {042509},
  year    = {2013},
  doi     = {10.1103/PhysRevA.88.042509},
  url     = {https://doi.org/10.1103/PhysRevA.88.042509}
}

@article{Viotti2021,
  author  = {Viotti, Ludmila and Lombardo, Fernando C. and Villar, Paula I.},
  title   = {Enhanced decoherence for a neutral particle sliding on a metallic surface in vacuum},
  journal = {Physical Review A},
  volume  = {103},
  pages   = {032809},
  year    = {2021},
  doi     = {10.1103/PhysRevA.103.032809},
  url     = {https://doi.org/10.1103/PhysRevA.103.032809}
}

@article{Farias2020,
  author  = {Far{\'{i}}as, M. Bel{\'{e}}n and Lombardo, Fernando C. and Soba, Alejandro and Villar, Paula I. and Decca, Ricardo S.},
  title   = {Towards detecting traces of non-contact quantum friction in the corrections of the accumulated geometric phase},
  journal = {npj Quantum Information},
  volume  = {6},
  pages   = {25},
  year    = {2020},
  doi     = {10.1038/s41534-020-0252-x},
  url     = {https://doi.org/10.1038/s41534-020-0252-x}
}

@article{Bienfait2019,
  author  = {Bienfait, A. and Satzinger, K. J. and Zhong, Y. P. and Chang, H.-S. and Chou, M.-H. and Conner, C. R. and Dumur, E. and Grebel, J. and Peairs, G. A. and Povey, R. G. and Cleland, A. N.},
  title   = {Phonon-mediated quantum state transfer and remote qubit entanglement},
  journal = {Science},
  volume  = {364},
  pages   = {368--371},
  year    = {2019},
  doi     = {10.1126/science.aaw8415},
  url     = {https://doi.org/10.1126/science.aaw8415}
}

@article{Brattegard2026,
  title = {Correlated decoherence and thermometry with mobile impurities in a one-dimensional Fermi gas},
  author = {Brattegard, Sindre and Fogarty, Thom\'as and Busch, Thomas and Mitchison, Mark T.},
  journal = {Phys. Rev. A},
  volume = {113},
  issue = {1},
  pages = {013302},
  numpages = {13},
  year = {2026},
  month = {Jan},
  publisher = {American Physical Society},
  doi = {10.1103/prvd-7xjr},
  url = {https://link.aps.org/doi/10.1103/prvd-7xjr}
}


\appendix

\section{Reduced spectral representation and Jacobian}
\label{reduced_formula}

We derive the reduced one-dimensional representation of the auxiliary in-out contribution ${\rm Im}\,\Gamma_{\rm cross}$ used in the main text and make explicit the Jacobian associated with the resonance constraint. Starting from Eqs.~\eqref{ImGamma_spectral_new}, \eqref{rho_pm}, and \eqref{boundary_dispersion}, one obtains
\begin{equation}
\begin{aligned}
{\rm Im}\,\Gamma_{\rm cross}
=&\frac{\lambda_A^2\lambda_B^2}{2}
\frac{\pi^2u_\phi^4}{(2\pi)^3}
\int d\omega\,dk_x\,dk_y\,
\frac{1}{\Omega_{\mathbf{k}}^2}\delta(\omega-\Omega_{\mathbf{k}})\\
&\times\delta(\omega-vk_x+\Omega_{\mathbf{k}})
\left|G_h^R(\omega,\mathbf{k};a,0)\right|^2.
\end{aligned}
\label{A_after_sub}
\end{equation}

Performing the $\omega$ integration yields
\begin{equation}
\begin{aligned}
{\rm Im}\,\Gamma_{\rm cross}
=&\frac{\lambda_A^2\lambda_B^2}{2}
\frac{\pi^2u_\phi^4}{(2\pi)^3}
\int dk_x\,dk_y\,
\frac{1}{\Omega_{\mathbf{k}}^2}\\
&\times\delta(vk_x-2\Omega_{\mathbf{k}})
\left|G_h^R(\Omega_{\mathbf{k}},\mathbf{k};a,0)\right|^2.
\end{aligned}
\label{A_after_omega}
\end{equation}

\begin{equation*}
\beta^2 \equiv k_y^2+m_\phi^2u_\phi^2,
\qquad
\Omega_{\mathbf{k}}=u_\phi\sqrt{k_x^2+\beta^2}.
\end{equation*}

The resonance condition $vk_x-2\Omega_{\mathbf{k}}=0$ reproduces the on-shell momentum in Eq.~\eqref{kxstar_new}, which exists only for $v>2u_\phi$. Using
\begin{equation*}
\delta(f(k_x))=\sum_i \frac{\delta(k_x-k_{x,i})}{|f'(k_{x,i})|},
\qquad
f(k_x)=vk_x-2u_\phi\sqrt{k_x^2+\beta^2},
\end{equation*}
one finds at $k_x=k_x^\ast$
\begin{equation*}
f'(k_x^\ast)=\frac{v^2-4u_\phi^2}{v},
\end{equation*}
and therefore
\begin{equation}
\delta(vk_x-2\Omega_{\mathbf{k}})=\frac{v}{v^2-4u_\phi^2}\delta(k_x-k_x^\ast).
\label{A_jacobian_delta}
\end{equation}

Substituting Eq.~\eqref{A_jacobian_delta} into Eq.~\eqref{A_after_omega} gives
\begin{equation}
\begin{aligned}
{\rm Im}\,\Gamma_{\rm cross}
=&\frac{\lambda_A^2\lambda_B^2}{2}\frac{\pi^2u_\phi^4}{(2\pi)^3}\int dk_y\frac{1}{\Omega_{\mathbf{k}_\ast}^2}\frac{v}{v^2-4u_\phi^2}\\
&\times\left|G_h^R(\omega_\ast,\mathbf{k}_\ast;a,0)\right|^2,
\end{aligned}
\label{A_after_kx}
\end{equation}
with $\mathbf{k}_\ast=(k_x^\ast,k_y)$ and $\omega_\ast=\Omega_{\mathbf{k}_\ast}$. Using
\begin{equation*}
\frac{1}{\Omega_{\mathbf{k}_\ast}^2}\frac{v}{v^2-4u_\phi^2}=\frac{1}{u_\phi^2\,v}\frac{1}{k_y^2+m_\phi^2u_\phi^2},
\end{equation*}
one recovers the reduced expression quoted in Eq.~\eqref{ImGamma_reduced_new}. The corresponding Jacobian factor is
\begin{equation*}
\mathcal{J}^{-1}=\left|\frac{d}{dk_x}\left(vk_x-2\Omega_{\mathbf{k}}\right)\right|^{-1}_{k_x=k_x^\ast}=\frac{v}{v^2-4u_\phi^2}.
\end{equation*}

\section{Symmetric-phase benchmark}
\label{symmetric_phase}

For comparison, we summarize the Gaussian description in the symmetric phase, where $\langle\psi\rangle=0$.

The quadratic bulk action is
\begin{equation}
\begin{aligned}
S_{\psi,{\rm quad}}
=&\int dt\,d^3x\,\Theta(z)\Theta(a-z)\psi^\ast\\
&\times\left[\frac{(\partial_t-i\mu)^2}{u_\psi^2}-\nabla^2-M^2\right]\psi.
\end{aligned}
\end{equation}

The bulk propagator satisfies
\begin{equation}
\left[-\partial_z^2+\tilde\kappa^2(\omega,\mathbf{k})\right]G_0(\omega,\mathbf{k};z,z')=\delta(z-z'),
\end{equation}
with
\begin{equation}
\tilde\kappa^2(\omega,\mathbf{k})=k_\parallel^2+M^2-\frac{(\omega-\mu)^2}{u_\psi^2}.
\end{equation}
In the evanescent regime,
\begin{equation}
G_0(\omega,\mathbf{k};z,z')\approx \frac{1}{2\tilde\kappa}\,e^{-\tilde\kappa|z-z'|}.
\end{equation}

The boundary interaction is
\begin{equation}
S_{\rm int}=\int dt\,d^2x_\parallel
\left[g_A\phi_A|\psi|^2\big|_{z=a}+g_B\phi_B|\psi|^2\big|_{z=0}\right].
\end{equation}

The dissipative contribution takes the same kinematic form as in the main text, with the condensed-phase kernel replaced by the symmetric-phase response,
\begin{equation}
\begin{aligned}
{\rm Im}\,\Gamma_{\rm cross}^{({\rm sym})}
=&\frac{g_A^2g_B^2}{2}\int\frac{d\omega\,d^2\mathbf{k}}{(2\pi)^3}\rho_A^{(-)}(\omega-vk_x,\mathbf{k})\\
&\times\rho_B^{(+)}(\omega,\mathbf{k})
\left|\Pi_0^R(\omega,\mathbf{k};a,0)\right|^2.
\end{aligned}
\end{equation}

Applying the same on-shell reduction that leads to Eqs.~\eqref{res_condition_new} and \eqref{ImGamma_reduced_new}, one obtains
\begin{equation}
{\rm Im}\,\Gamma_{\rm cross}^{({\rm sym})}=\frac{g_A^2g_B^2u_\phi^2}{8\pi v}
\int_{-\infty}^{\infty}\frac{dk_y}{k_y^2+m_\phi^2u_\phi^2}\left|\Pi_0^R(\omega_\ast,\mathbf{k}_\ast;a,0)\right|^2.
\end{equation}
The threshold condition is again $v>2u_\phi$.


\end{document}